\documentclass{article}

\usepackage[preprint]{jmlr2e}
\usepackage{mathtools}

\usepackage{amsmath,amsfonts,bm}

\def\eqref#1{equation~\ref{#1}}

\def\1{\bm{1}}

\DeclareMathAlphabet{\mathsfit}{\encodingdefault}{\sfdefault}{m}{sl}
\SetMathAlphabet{\mathsfit}{bold}{\encodingdefault}{\sfdefault}{bx}{n}

\usepackage{hyperref}
\usepackage{url}
\usepackage{booktabs}
\usepackage{tabularx}
\usepackage[table]{xcolor}
\usepackage{colortbl}
\usepackage{makecell}
\usepackage{enumitem}
\usepackage{wrapfig}
\usepackage{titletoc}

\usepackage{lastpage}

\ShortHeadings{Accurate MI Estimation in High Dimensional Data}{Abdelaleem, Martini and Nemenman}
\firstpageno{1}

\title{Accurate Estimation of Mutual Information in High Dimensional Data}

\author{%
  \name Eslam Abdelaleem \thanks{Equal contribution}
        ~\thanks{Currently at Georgia Institute of Technology, Schools of Physics \& Psychology
        (\texttt{eslam.abdelaleem@gatech.edu}).}
        \email eslam.abdelaleem@emory.edu \\
  \addr Department of Physics \\
        Emory University \\
        Atlanta, GA 30322, USA
  \AND
  \name K.~Michael Martini \footnotemark[1]
        \email karl.michael.martini@emory.edu \\
  \addr Department of Physics \\
        Emory University \\
        Atlanta, GA 30322, USA
  \AND
  \name Ilya Nemenman \thanks{Corresponding author}
        \email ilya.nemenman@emory.edu \\
  \addr Departments of Physics and Biology \\
        Emory University \\
        Atlanta, GA 30322, USA
}

\editor{}

\begin{document}

\maketitle

\begin{abstract}

Mutual information (MI) quantifies statistical dependence between variables and is widely used across scientific disciplines, yet accurate estimation from finite data remains notoriously difficult. No estimator is universally accurate, and common approaches fail in the high-dimensional, undersampled regimes typical of modern experiments, where the number of samples $N$ is comparable to or smaller than the data dimensionality $K$. Neural network--based estimators offer promise in these regimes, but their accuracy is sensitive to the  dataset size, latent structure, and hyperparameters. Crucially, no accepted tests exist to detect when they fail, making them effectively unusable in scientific applications.

We show that neural network MI estimators can be made reliable when the statistical dependencies in the underlying joint probability distribution admit a low-dimensional latent representation. As we confirm numerically and argue analytically, in this case, sample complexity is governed by the latent dimensionality $K_Z \ll K$ rather than the ambient dimension. Building on this insight, we develop a practical protocol that improves typical neural MI estimators by providing them with features they do not usually possess: explicit statistical consistency checks, bias correction, and confidence intervals. We additionally introduce a new class of probabilistic critics (the Variational Symmetric Information Bottleneck, or VSIB, family) that substantially reduce bias and variance at higher MI values where standard estimators break down.

We validate the protocol across a range of examples. On synthetic benchmarks, we demonstrate reliable estimation with $K=500$ and $N$ as low as $256$. We show that our protocol correctly handles the full 40-dataset standard suite of test cases by \citet{czyz2024beyond}, matching or exceeding standard stand-alone estimators while being the only method to report confidence intervals and flag unreliable estimates. On real-image data, the protocol succeeds on noisy MNIST ($K=784$) and on CIFAR-10/100 ($K=3072$) with a ResNet-20 backbone, where we achieve reliable MI detection for datasets sizes well below the ambient pixel dimension. Together, these results increase the range of applicability of neural MI estimation and clarify the conditions under which it can be trusted.
\end{abstract}

\section{Introduction}
\label{intro}

Mutual information (MI) is a fundamental measure of statistical dependence between two variables \citep{shannon1948mathematical}. It captures both linear and nonlinear associations, is invariant under reparameterizations, and is zero if and only if the variables are statistically independent. This makes MI a key tool across disciplines, from neuroscience to computer vision \citep{viola1997alignment}. In systems neuroscience, MI estimation plays an important role in understanding neural coding, analyzing spike trains in single neurons and neural populations, and studying patterns of information transfer across brain areas and behaviors \citep{tang2014millisecond, palmer2015predictive, panzeri2001role, strong1998entropy, pica2017quantifying, runyan2017distinct, pascual2024millisecond}. Similarly, in brain imaging, MI quantifies functional connectivity between brain regions, illuminating  effects of neurological disorders \citep{ince2017statistical, hlinka2011functional, li2022functional}.  MI has also proven useful in domains like protein sequence alignment and contact prediction \citep{marks2011protein, lu2020self, sgarbossa2023generative}, and in the inference of gene regulatory networks \citep{margolin2006aracne}. In computer vision, state-of-the-art models, such as CLIP \citep{radford2021learning}, utilize loss functions related to MI to align visual and textual representations. Similarly, self-supervised learning frameworks such as Barlow Twins \citep{zbontar2021} can be interpreted as mutual information-based objectives designed to enforce cross-modal correspondences \citep{abdelaleem2025deep}.

For continuous variables $X$ and $Y$, MI is 
$I(X;Y) = \int dx\ dy\, p(x,y) \log_2  \frac{p(x,y)}{p(x)p(y)}$ (measured in \emph{bits}\footnote{Most estimators are naturally expressed in \emph{nats}, where $\log$ is the natural logarithm. Thus in all {\em derivations} in this paper $\log=\ln$. However, when reporting the MI values, we convert to \emph{bits} for clarity, $\log=\log_2$.}), 
where $x$ and $y$ are specific values of the variables, $p(\cdot)$ are the corresponding probability densities, and the integration is over the domain of the variables. (For discrete $X$ and $Y$, sums over distributions are used instead.) Since MI is a nonlinear function of  $p(x,y)$, substituting an unbiased estimate of $p$ into the definition of MI results in a biased estimate of MI. Typically, the bias of estimators is a more serious problem than their variance, particularly for continuous variables, because MI is invariant under reparameterization, while it is impossible to construct an estimator that is covariant under all reparameterizations \citep{holy2002impossibility}. Traditional methods, such as histogram-based, k-nearest neighbors (kNN), box-counting, or kernel-based \citep{kraskov2004estimating, gao2015efficient, ross2014mutual, amir2014estimation,khan2007relative,steuer2002mutual,daub2004estimating,trappenberg2005input,fransens2004multimodal}, struggle to reduce the bias for high-dimensional data since they require the number of samples that grows exponentially with the dimensionality \citep{walters2009estimation,khan2007relative,gao2015efficient,czyz2024beyond}. 

Recent advances in machine learning resulted in  neural network (NN)-based estimators for MI, which aim to circumvent the limitations of traditional methods. These estimators frame MI estimation as optimization over a family of functions \citep{barber2003information, barber2004algorithm, nguyen2010estimating, poole2019variational, song2019understanding, oord2018representation}. In principle, they can work even for very high-dimensional data that are out of reach for traditional methods. For instance, they supposedly  compute MI between images, with the dimensionality of thousands. However, the practical accuracy of these methods remains unclear.  First, most of them have been tested primarily on synthetic data with simple dependence structures and unrealistically large datasets. Second, since universally good MI estimation without smoothness assumptions on the underlying distribution is impossible (see, e.g., \citet{paninski2003estimation,kandasamy2015nonparametric}), internal consistency checks are essential to signal whether the output can be trusted. Such checks are not widely adopted. Third, NN estimators depend on hyperparameters, such as criteria for stopping training, and optimal parameter choices are unclear. 

Here, we systematically address these gaps. We argue that successful neural MI estimation requires: (i)~the data having a \emph{low-dimensional latent structure}, even if the observed dimensionality is high; (ii)~the critic being \emph{sufficiently expressive}---in particular, having an embedding dimension matched to the latent space; and (iii)~sufficient data to \emph{resolve statistical dependencies in the latent}, not the full data space. Our specific contributions are: (1)~a practical, estimator-agnostic \emph{protocol} for reliable MI estimation, consisting of an early-stopping heuristic, internal bias checks via subsampling and extrapolation, and confidence intervals---the latter being particularly rare among MI estimators; (2)~a new class of \emph{probabilistic critics} (the Variational Symmetric Information Bottleneck, or VSIB, family) that substantially reduce bias and variance at high MI values where standard estimators break down; and (3)~a systematic \emph{benchmarking} across classical and recent neural estimators on synthetic and real-world data, demonstrating
when and why MI estimation succeeds and when it fails.

\paragraph*{Paper organization.}
Section~\ref{sec:background} reviews traditional and neural MI estimators, including recent approaches, and identifies the key limitations our work addresses. Section~\ref{sec:unifying_critics} introduces the generalized critic framework and the VSIB probabilistic critic family. Section~\ref{txt:results} presents our main results: estimator behavior in the infinite- and finite-data regimes, the role of latent dimensionality, and the complete estimation protocol with validation on synthetic benchmarks, noisy MNIST, and CIFAR-10/100. Section~\ref{sec:discussion} discusses implications and limitations. Theoretical derivations, protocol details, extended evaluations, and implementation specifications appear in Appendices~\ref{app:theory}, \ref{app:protocol}, \ref{app:evaluations}, and~\ref{app:implementation}, respectively.

\section{Background}
\label{sec:background}

\paragraph*{Traditional MI Estimation.}
Estimating mutual information (MI) from finite data is notoriously challenging, especially for high-dimensional continuous variables \citep{paninski2003estimation}. To see this, consider a simple argument: suppose each component of $X$ and $Y$ lies within a bounded range of size $A$, and the joint density $p(x, y)$ is smooth, with its smallest feature on the scale of $a$. Then accurate estimation requires  $N \gg (A/a)^{K_{\text{tot}}}$ samples, where $K_\text{tot} = K_X + K_Y$ is the total dimensionality of $X$ and $Y$. If $A/a>1$, the required sample size is exponential in $K_\text{tot}$, illustrating the classic curse of dimensionality. The situation is even worse when the coordinate system in which $p$ is smooth is unknown \citep{holy2002impossibility}, or when the variables are unbounded.
Thus, while many methods have been developed to estimate MI, they often break down beyond $K_{\rm tot} \sim 10$ dimensions \cite{holmes2019estimation}. In contrast, most modern datasets are high-dimensional, e.g., images with thousands of pixels, or neural recordings from thousands of units.

\paragraph*{Neural Network-Based Estimators.}
The struggle against the curse of dimensionality has motivated neural network (NN)-based approaches since deep NNs can capture complex nonlinear dependencies in high-dimensional data \citep{lecun2015deep}. Neural variational methods have become particularly influential for MI estimation \citep{barber2003information, barber2004algorithm, nguyen2010estimating, donsker1983asymptotic}. Typically, we do not have access to the full joint distribution $p(x, y)$ or the marginals $p(x)$ and $p(y)$, but we can draw samples from them. Variational estimators leverage this by reformulating MI in terms of a Kullback-Leibler (KL) divergence:
\begin{equation}
I(X;Y) = D_{\text{KL}}\left(p(x,y) \parallel p(x)p(y)\right) = 
\mathbb{E}_{p(x)}\left[D_{\text{KL}}\big(p(y|x) \parallel p(y)\big)\right].
\label{eq:mi_dkl_2}
\end{equation}
Then using the Donsker–Varadhan (DV) representation of the first KL divergence in Eq.~(\ref{eq:mi_dkl_2}) \citep{donsker1983asymptotic}, $D_{\text{KL}}(P \parallel Q) \ge \max_T \left\{\mathbb{E}_P[T] - \log \mathbb{E}_Q[e^{T}]\right\}$, learning the {\em critic} function $T(x, y)$ via a NN, and replacing expectations with sample averages, one obtains the MINE estimator \citep{Belghazi2018MutualEstimation, poole2019variational} (Eq.~\ref{eq:mine2}). However, MINE suffers from high variance, and its estimate is not a strict lower bound on MI when the normalization term is approximated by Monte Carlo sampling \citep{poole2019variational}. One addresses this by clipping the critic to the range $\pm\tau$, yielding the SMILE estimator \citep{song2019understanding} (Eq.~\ref{eq:smile2}). Alternatively, applying the DV representation to the second KL divergence in Eq.~(\ref{eq:mi_dkl_2}) instead leads to the InfoNCE estimator \citep{oord2018representation}, widely used in contrastive learning (Eq.~\ref{eq:infonce2}). See derivations for all the methods in Appx.~\ref{app:dv_derivations}.
Although early studies employed simple critic networks, the architecture and expressivity of $T(x,y)$ strongly influence estimator performance. We return to this in Sec.~\ref{sec:unifying_critics}.

\paragraph*{Recent Extensions and Alternatives.}
Several recent approaches have attempted to address  limitations of foundational DV estimators through increased model expressivity or clever parameterizations. \emph{Generative estimators} such as MINDE \citep{franzese2024minde} reframe MI estimation using score-based diffusion models, estimating KL divergences from learned score functions rather than discriminative critics. This sidesteps the partition function variance of variational lower bounds (VLB) estimators but requires computationally and data demanding training of full diffusion models. \emph{$f$-divergence estimators} such as f-DIME \citep{letizia2024mutual} decouple the training objective from the final MI estimate: a VLB loss is used only to learn the optimal critic, and MI is then estimated directly from the learned density ratio, avoiding the partition function entirely. \emph{Normalizing flow approaches} \citep{butakov2024mutual} learn invertible dimension-preserving transformations that map the data to a canonical Gaussian space, after which MI reduces to a closed-form sum over canonical correlations. Crucially, because these transformations are bijective rather than compressive, they do not reduce dimensionality and thus inherit the same sample complexity as operating in the full ambient space. \emph{Latent compression methods} such as LMI \citep{gowri2024approximating} first reduce dimensionality of $X$ and $Y$ independently rather than simultaneously and then estimate MI in the compressed space, an approach that was shown to overlook the shared signal in some cases \citep{abdelaleem2025deep, abdelaleem2024simultaneous}. Overall, these approaches address known limitations of standard DV-based estimators, but  usually at the cost of substantially greater sample and computational complexity, which is the wrong trade-off for the high-dimensional, undersampled regime that we focus on. As we demonstrate in Appx.~\ref{app:evaluations}, these methods either fail or produce unreliable results when $N \lesssim K$, even on controlled benchmarks where ground truth is known. Crucially, none of them provide a mechanism to detect when their estimates are untrustworthy, which is a central problem that we address.

\paragraph*{Limitations of Neural Estimators.}
Despite their popularity, it is unclear if existing neural MI estimators are truly accurate, calling their widespread use into question. First, most tests \citep{czyz2024beyond} of the estimators to date have been performed on synthetic data with low dimensionality (e.g., $K_X, K_Y \sim 10$), where the true MI is known analytically. However, traditional estimators such as kNN-based methods \citep{kraskov2004estimating, holmes2019estimation} already perform well then \citep{czyz2024beyond}. Unless neural estimators clearly outperform these simpler methods in high dimensions ($K \gtrsim 100$), their practical value is limited. Yet, evaluations in this regime remain scarce.

Second, when $X$ and $Y$ are jointly Gaussian, MI can be computed exactly from their correlation matrix. Since correlation matrices can be reliably estimated when $K/N \ll 1$ \citep{bouchaud2007large,swain2025distribution}, this provides a natural benchmark. If a neural estimator fails when a linear method succeeds, it is not exploiting all the statistical structure in the data. Nevertheless, such comparisons are rare. As we will demonstrate below, some neural estimations fall short on this metric.

Finally, estimators are often validated in effectively infinite-sample regimes \citep{poole2019variational, song2019understanding}, with a fresh data batch at every training step. This sidesteps overfitting and gives an overly optimistic view of estimator performance. In real life, sample size is often small ($N \sim K$), and success for infinite-data does not imply practical utility. We will show that unbiased estimation is sometimes possible even in this heavily undersampled regime if data have a simple latent structure.

\section{A Generalized Critic}
\label{sec:unifying_critics}

Neural network-based MI estimators typically rely on a \emph{critic} function $T(x, y)$ that approximates a log-density ratio. In a unified formulation, the critic can be expressed as:
\begin{equation}
\label{eq:generalized_critic}
T(x, y) = f\left(g\left(x\right), h\left(y\right)\right),
\end{equation}
where $g: \mathcal{X} \to \mathcal{Z}_X$ and $h: \mathcal{Y} \to \mathcal{Z}_Y$ are embedding functions, and $f: \mathcal{Z}_X \times \mathcal{Z}_Y \to \mathbb{R}$ combines the embeddings into a scalar score. Different choices of $f$, $g$, and $h$ recover many well-known estimators and permit novel architectures tailored to specific tasks. Specifically:

\textbf{Joint (Concatenated) Critic:} Setting $g$, $h$ to the identity maps and letting $f$ be a NN that operates on the concatenated inputs reproduces the joint-critic architecture used in MINE~\citep{Belghazi2018MutualEstimation}.

\textbf{(Deep) Separable Critic:} Choosing $f(g, h) = \langle g, h \rangle$, with $g$, $h$ being vector-valued embeddings (e.g., via multilayer perceptrons), yields the separable critic of  InfoNCE~\citep{oord2018representation}.

Although certain critics are commonly paired with specific objectives, this pairing is not obligatory. For instance, separable critics, typically used with contrastive losses, can be combined with non‑contrastive objectives. Likewise, joint critics can be used in contrastive settings.

In this work, we introduce or reformulate additional critic choices:

\textbf{Concatenated Quadratic Critic:} We show that if $g$ and $h$ are linear projections (e.g., identity maps) and $f$ is a quadratic form of its concatenated arguments, then MI estimation, denoted here as $I_{\rm CCA}(X;Y)$, reduces to measuring canonical correlations \citep{Hotelling1936} in a shared low-dimensional space (Appx.~\ref{app:cca_derivation}).

\textbf{Probabilistic Critic (Variational Symmetric Information Bottleneck, VSIB, Framework):} Rather than using deterministic mappings, $f$, $g$, and $h$ can be stochastic, leading to variational objectives that regularize the embedding distributions. We implement this with a loss similar to that introduced by \citet{abdelaleem2025deep} (Appx.~\ref{app:vsib}):
\begin{equation}
\label{eq:vsib}
L_{\text{EST}-\text{VSIB}} = I^{E}(X; Z_X) + I^{E}(Y; Z_Y) - \beta I^{D}_{\text{EST}}(Z_X; Z_Y),
\end{equation}
where the $I^E$ terms are KL-divergence penalties that encourage the stochastic encoders to produce compact, well-regularized embeddings, and $I^D_{\text{EST}}$ measures the mutual information between the latent representations using any chosen neural estimator.

The regularization provided by the $I^E$ terms prevents overfitting by constraining the embeddings $Z_X$ and $Z_Y$ to remain close to a standard Gaussian prior. This acts as an intrinsic regularizer on the critic's capacity: it cannot form arbitrarily sharp, sample-specific embeddings. As Fig.~\ref{fig:high_dim_finite_vs_samples} will show, the benefit is most pronounced in high MI regimes where resolving fine-grained structure requires large $N$, which in turn would lead to overfitting if not for the regularization.

\section{Results}
\label{txt:results}
\subsection{Estimator Performance in the Infinite Data Regime}
We first evaluate neural MI estimators in an idealized, effectively infinite-sample setting that eliminates overfitting. Although common in prior work, this regime obscures many real-world challenges. Thus we use it here only to isolate estimator behavior from finite-sample effects before introducing them in Sec.~\ref{txt:performanceFinite}.

Our synthetic benchmarks follow a progression of increasing difficulty, mirroring standard practice in MI estimation research~\citep{czyz2024beyond,song2019understanding}. We begin with jointly Gaussian data, where MI has a closed-form solution and a bilinear critic suffices. We then introduce monotone nonlinearity (cubing one variable), which breaks linearity while preserving invertibility. Finally, we pass both variables through separate randomly initialized teacher networks, creating strongly nonlinear, partially non-invertible
dependencies. This progression isolates the contribution of nonlinearity to estimator difficulty and identifies the regimes where neural estimators offer genuine advantages over classical methods. We then extend each setting to high observed dimensionality ($K=500$) while keeping the latent dimension fixed ($K_Z=10$), separating the effects of ambient and latent dimensionality.

We adopt the following notation. $X$ and $Y$ are the variables whose MI we estimate; $Z_X$ and $Z_Y$ denote their low-dimensional embeddings by a critic. The observed dimensionalities are $K_X$ and $K_Y$ ($K$ if both are equal), while $K_Z$ is the true latent dimensionality when it differs from $K$. Finally, $k_Z$ is the embedding dimensionality of the critic, which may differ from $K_Z$.

\paragraph*{Low-Dimensional \texorpdfstring{$X$ and $Y$}{X and Y}.}

\begin{figure}[tbp]
    \includegraphics[width=\textwidth]{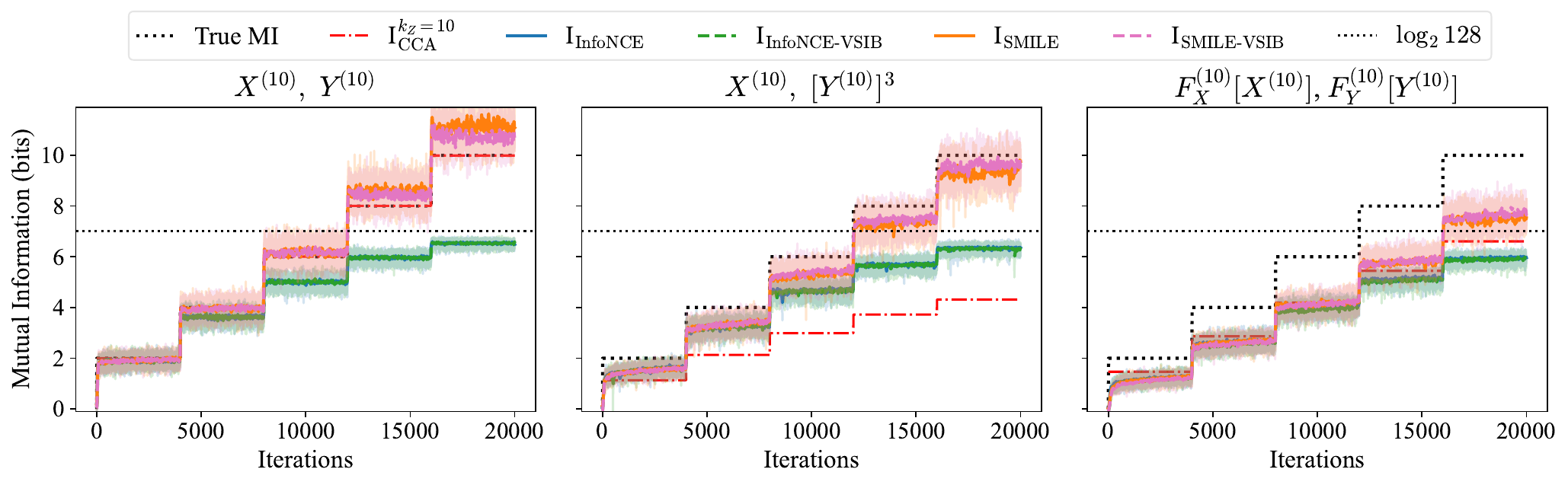}
    \caption{\textbf{MI estimators in the low-dimensional, infinite-data regime.} Each panel plots running MI estimates over training iterations for five true MI levels (increasing every 4000 iterations). Each step introduces a fresh batch of 128 samples. We compare the CCA-based estimator (optimal for Gaussian data), InfoNCE, SMILE, and their probabilistic variants (denoted with VSIB). Faint curves show raw estimates; bold curves show smoothed trends (Appx.~\ref{app:implementation}). {Left:} For jointly Gaussian $X, Y$, all estimators initially perform well. InfoNCE plateaus at its well-known intrinsic upper bound \citep{oord2018representation} $\log$(batch size) $\approx 7$ bits, while SMILE begins to overestimate at high MI, indicating overfitting. $I_{\rm CCA}$  overlaps with ground truth, as expected. {Middle:} Cubing $Y$ breaks linearity, and  $I_{\rm CCA}$ fails. Nonetheless, InfoNCE behavior is almost unchanged, and SMILE remains reasonably effective with sufficient training (Appx.~Fig.~\ref{si_fig:low_dim_infinite_more_epochs}). Both slightly underestimate at low MI, and for SMILE this is largely offset by its intrinsic positive bias at high MI.  {Right:} Passing $X$ and $Y$ through separate frozen teacher networks (one hidden softplus layer, 1024 units) creates highly nonlinear dependencies. All estimators  underestimate MI.}
    \label{fig:low_dim_infinite_main}
\end{figure}

We begin with low-dimensional data ($K_X = K_Y = 10$), where both traditional (e.g., kNN) and neural estimators are expected to perform well. We first analyze jointly Gaussian data with a chosen correlation. It is well known that, in this case, MI estimation reduces to the sum of information in each canonical correlation pair \citep{gelfand1959calculation,kullback1997information,huffmann2022distribution}, $I_{\rm CCA}=-\frac{1}{2}\sum_{i}^{K_Z}\log(1-\rho_i^2)$, where $\rho_i$ is the canonical correlation. We then increase complexity by (i) cubing $Y$ or (ii) passing both variables through separate fixed teacher networks, thereby introducing diverse nonlinear dependencies. 

Figure~\ref{fig:low_dim_infinite_main} (left) confirms that InfoNCE, SMILE, and CCA track the true MI  on Gaussian data (until InfoNCE saturates \citep{oord2018representation}). The data matrices used here are aggregated across all batches seen during training at a given MI level.  If the critic is expressive enough, SMILE is expected to show high variance and to overestimate at high MI since it evaluates log-sum-exp in its DV bound, Eq.~(\ref{eq:mine2}), which is biased due to the Jensen's inequality \citep{guo2022tight,choi2020regularized}. We observe both problems, but neither is strong, at least at low MI. 

Cubing $Y$ introduces nonlinear correlations, and $I_{\rm CCA}$ fails, Fig.~\ref{fig:low_dim_infinite_main} (middle). In contrast, neural estimators recover MI after sufficient training (Appx.~Fig.~\ref{si_fig:low_dim_infinite_more_epochs}). Both exhibit a small but significant negative bias at low MI, likely due to the reduced invertibility of the cubic map near $y=0$.
When $X$ and $Y$ are passed through separate frozen teacher networks, Fig.~\ref{fig:low_dim_infinite_main} (right), all estimators underestimate MI, although neural methods still outperform the CCA baseline. We again attribute this bias to the reduced invertibility, now due to the softplus saturation in the teacher networks.

Appx.~Fig.~\ref{si_fig:low_dim_infinite_all} surveys other neural estimators and confirms that  InfoNCE and SMILE consistently outperform the rest. Therefore, in what follows, we restrict our analysis to these two methods, using a fixed clipping factor $\tau=5$ for SMILE. Combined, these results validate MI neural estimators---specifically InfoNCE, SMILE, and their VSIB variants---in low dimensions with abundant data. However, they still underestimate MI for strongly nonlinear dependencies, even with effectively unlimited data. Thus, seemingly, there is little reason to prefer them over simpler correlation‑based or kNN methods, which already perform well in these scenarios  \citep{kraskov2004estimating,holmes2019estimation,czyz2024beyond}.

\begin{figure}[tbp]
    \centering
    \includegraphics[width=\textwidth]{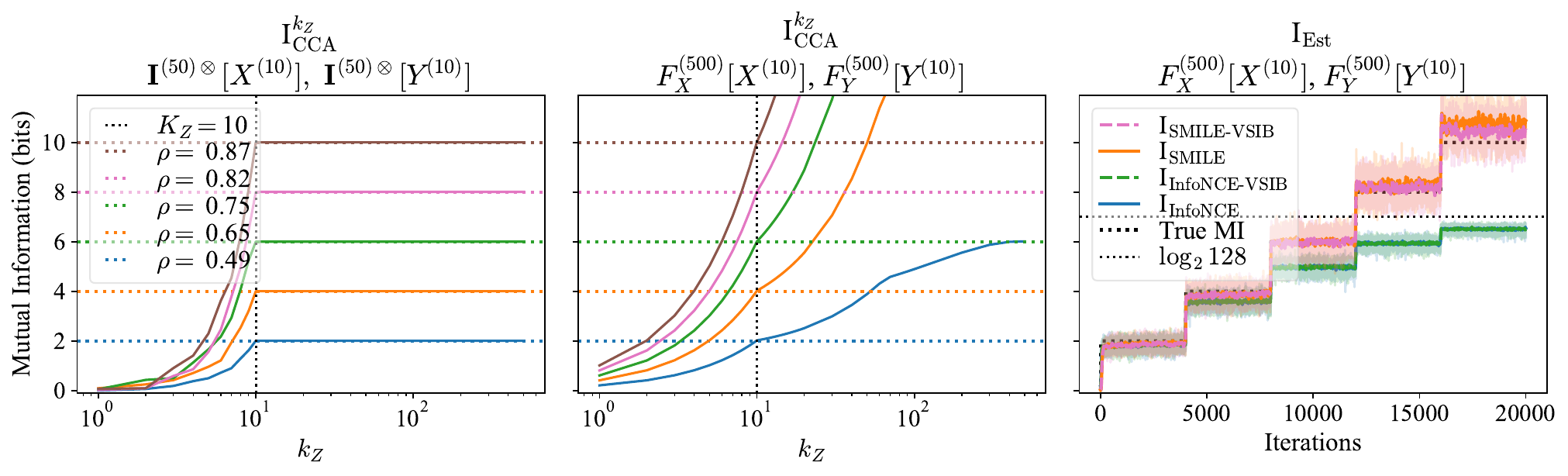}
    \caption{\textbf{MI estimators in the high-dimensional, infinite-data regime.} We extend Fig.~\ref{fig:low_dim_infinite_main} to $K_X = K_Y = 500$, embedding $K_Z = 10$ latent variables into high-dimensional $X$ and $Y$.  {Left:} A linear transformation (e.g., identity and replication) expands $Z$ to 500-dimensional $X$ and $Y$. $I_{\rm CCA}$ with $k_Z\ge K_Z=10$ dimensions still accurately recovers the ground truth. {Middle:} A frozen nonlinear teacher network maps $Z$ to 500-dimensional $X$ and $Y$.  Unlike the linear case, $I_{\rm CCA}$ fails due to the nonlinearity of the transformation. Increasing $k_Z>K_Z$ detects spurious correlations, inflating MI estimates and illustrating the limitations of linear methods in nonlinear settings. {Right:} Neural estimators (InfoNCE, SMILE, and their VSIB variants) are applied directly to the full 500-dimensional data. All are accurate across the full range of true MI values, performing even better than in Fig.~\ref{fig:low_dim_infinite_main} due to improved invertibility of the nonlinear transformation in high dimensions.}
    \label{fig:high_dim_infinite_main}
\end{figure}

\paragraph*{High-Dimensional \texorpdfstring{$X$, $Y$}{X and Y}.}
To distinguish effects of the observed and latent dimensionality, we increase the former, while fixing the latter. We start again with the jointly Gaussian 10-d $X$ and $Y$ with the known ground truth MI. We then (i) replicate each of the ten components of $X$ and $Y$ 50 times (denoted $X\to {\mathbf I}^{(50)\otimes}[X]$), and (ii) pass $X$, $Y$ through distinct frozen teacher networks, embedding each into 500 dimensions (denoted $X\to F_X^{(500)}[X]$, and similarly for $Y$). Both cases result in $K_X=K_Y\equiv K=500$, while $K_Z=10$, with the former case having only linear correlations.

Figure \ref{fig:high_dim_infinite_main} shows that, in the linear replication setup, $I_{\rm CCA}$  accurately recovers the true MI when $k_Z\ge K_Z$.  In the nonlinear case, the CCA approach breaks down: as the number of detected canonical pairs $k_Z$ increases, the method approximates nonlinear dependencies with an ever larger set of linear projections, inflating the MI estimate with no upper bound. In contrast, neural estimators (InfoNCE, SMILE, and their VSIB variants) recover the ground truth MI when applied directly to the full $K=500$-dimensional spaces (up to  InfoNCE's saturation). Performance even surpasses the low-dimensional case, Fig.~\ref{fig:low_dim_infinite_main}, because non-invertible, saturated softplus regions in one random embedding can be inverted in others, allowing reconstruction of the full latent manifold.

This highlights two points. First, neural estimators may work well in high-dimensional nonlinear settings, where traditional approaches fail \citep{holmes2019estimation} (Appx.~Fig.~\ref{si_fig:KSG_examples}). Second, matching the estimator to the  structure of the data (e.g., using $I_{\rm CCA}$ for linear correlations) may result in more computationally and data-efficient estimation \citep{abdelaleem2024simultaneous}.

\subsection{Estimator Performance with Finite Datasets}
\label{txt:performanceFinite}

For  infinite-data,  neural estimators receive a fresh data batch at every training step. Yet, in practice, datasets are always finite. The impact of a finite sample size on the estimators remains poorly understood (though see \cite{czyz2024beyond}); here we address this gap.

\begin{figure}[t]
\centering
    \includegraphics[width=0.7\linewidth]{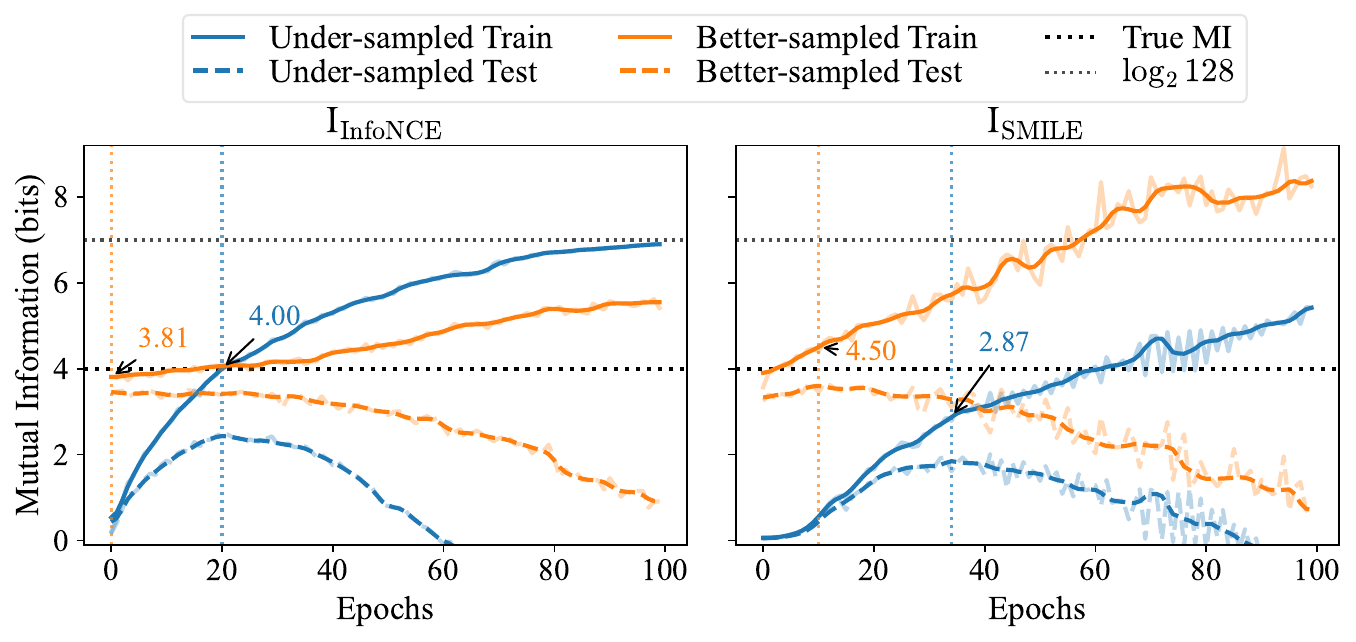}
    \caption{\textbf{The stopping heuristic.} We evaluate neural MI estimators for  finite-data using the teacher model from Fig.~\ref{fig:high_dim_infinite_main}, where 10 latent variables carrying 4 bits of MI are embedded in 500-dimensional $X$ and $Y$. We compare two sampling regimes for InfoNCE (left) and SMILE (right): 256 samples (\emph{under-sampled}) and a larger dataset of $2^{14} = 16{,}384$ samples (\emph{better-sampled}). In all cases, the test-set MI initially rises before declining due to overfitting (we do not show the negative values). The \textbf{stopping heuristic} selects the epoch with the peak test MI but reports  the corresponding training MI. Here the batch size is 128, so that InfoNCE does not saturate.}
    \label{fig:high_dim_finite_stop}
\end{figure}

\paragraph*{The Stopping Heuristic.}

Figure~\ref{fig:high_dim_finite_stop} demonstrates overfitting for finite data in the high-dimensional teacher model of Fig.~\ref{fig:high_dim_infinite_main} (ground-truth MI of 4 bits).  We track $I_{\rm InfoNCE}$ and $I_{\rm SMILE}$ on  training and held‑out sets for $2^8=256$ ({\em undersampled}) and $2^{14}=16,384$ ({\em better sampled})  pairs of  $K=500$-dimensional $X$ and $Y$ as training progresses. Because 64 epochs in the undersampled setting expose the network to the same number of examples as a single epoch in the better-sampled case (but with repetition), direct epoch counts are not comparable. All training curves start below the true MI and then rise (InfoNCE until the saturation) as the networks fit finer‑scale structure of the data distribution. Because the training curves show no clear inflection when they surpass the ground‑truth, common heuristics (fixed epoch counts or loss plateau) give no reliable stopping signal. In contrast, the test curves grow initially but soon collapse, revealing overfitting.

\begin{figure}[t] 
    \centering
    \includegraphics[width=0.7\linewidth]{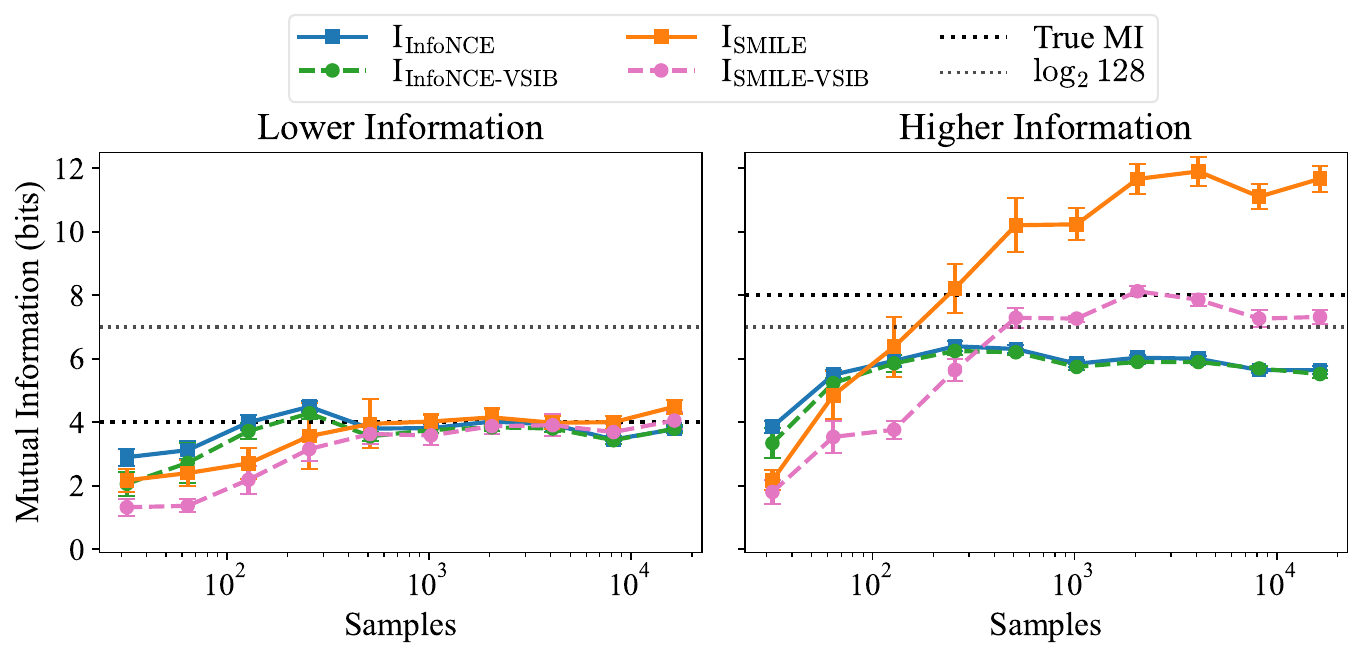}
    \caption{\textbf{MI vs.\ sample size: moderate MI (4 bits) and higher MI (8 bits).} We compare InfoNCE, SMILE, and $\text{VSIB}$ versions with the max-test stopping for different sample sizes. Data  from the frozen teacher model ($10$ latent, $500$ data dimensions). All estimators use separable critics, $k_Z=32$. Means $\pm$  s.d.\ over 10 trials shown. {Left:} For small MI (4 bits), all estimators recover the ground truth for $10^2\lesssim N< K=500$. {Right:} For high  MI (8 bits), contrastive estimators (InfoNCE and $\text{InfoNCE}-\text{VSIB}$) saturate near $\log(\text{batch size})=7$ bits. SMILE overestimates dramatically as $N$ grows. $\text{SMILE}-\text{VSIB}$ tracks the ground truth accurately for all $N\gtrsim 10^2$.}
    \label{fig:high_dim_finite_vs_samples}
\end{figure}

To devise a better stopping rule, note that the training time effectively sets the smallest scale resolved in the  neural approximation of the DV-optimal critic. Early stopping oversmooths and underestimates MI, whereas late stopping undersmooths, thus pushing the training MI high while the test MI falls. This mirrors kernel density estimation (KDE) of MI \citep{moon1995estimation}. The optimal resolution must depend on data complexity and sample size, so no fixed epoch rule would work.

In KDE MI estimation, one chooses the bandwidth that maximizes held-out likelihood \citep{margolin2006aracne}. Since the joint-density term dominates test error, this maximizes test MI. Analogously,  our stopping heuristic is: track  MI on a test batch each epoch, pick the epoch where this test value peaks, and report the corresponding \emph{training MI} as the estimate (see Appx.~\ref{app:explain_max_test_heuristic} for  justification). To our knowledge, such a rule has not been formalized for neural MI estimation.

\paragraph*{Probabilistic Embeddings Reduce Estimator Bias and Variance.}

Figure~\ref{fig:high_dim_finite_vs_samples} compares InfoNCE, SMILE, and their probabilistic variants for varying sample sizes and different ground truth MI for the $K=500$, $K_Z=10$ random teacher networks model. For moderate  MI (4 bits), all estimators converge to truth when $K_Z\ll N\allowbreak \sim O(10^2)<K$. This argues that MI estimation requires good sampling of the latent space,  but not necessarily the data space. As before, InfoNCE has lower variance than SMILE.

For higher true MI (8 bits), all variational bounds and corresponding estimators degrade \citep{poole2019variational}. InfoNCE and $\text{InfoNCE}-\text{VSIB}$ both saturate near $\log(\text{batch size})=7$ bits, as expected \citep{oord2018representation}.  Consistent with Figs.~\ref{fig:low_dim_infinite_main}, \ref{fig:high_dim_infinite_main},  SMILE substantially overestimates when large sample size allows overtraining. In contrast, $\text{SMILE}-\text{VSIB}$ remains accurate and stable, converging to the correct value at $K_Z\ll N\lesssim K_X+K_Y$. This confirms the utility of the VSIB family introduced in Sec.~\ref{sec:unifying_critics} showing that probabilistic regularization prevents the pathological overfitting that affects SMILE at higher MI.

\paragraph*{Low-dimensional Latent Structure Allows Reliable Estimation.} Our experiments have used data with statistical dependencies in a relatively low-dimensional latent space, $K_Z=10$. We now ask how the latent dimensionality limits neural MI estimation by simultaneously varying the true latent dimension, $K_Z$, and the critic’s embedding dimension, $k_Z$, while keeping the data dimension $K=500$ large  and fixed. We consider a low-dimensional latent setting, $K_Z=1$, a moderate regime, $1 \ll K_Z=100 \ll K$, and a fully high-dimensional latent regime, $1\ll K_Z=K=500$.  With the ground-truth MI at 4 bits (where all estimators can work, Fig.~\ref{fig:high_dim_finite_vs_samples}), each canonical pair contributes $0.4$, $0.04$, and $0.008$ bits  (equivalent correlation $\rho\approx 0.65$, $0.23$, and $0.11$, respectively), before they are nonlinearly mixed by the teacher networks. 
We expect that, if $k_Z\ge K_Z$, the critics can recover dependencies in the low-dimensional case (similar to \cite{abdelaleem2024simultaneous}). However, as $K_Z$ increases, the critic must disentangle an ever-growing number of weak interactions, and its MI estimate will deteriorate at a fixed $N$.

\begin{figure}[!t]
\centering
\includegraphics[width=\textwidth]{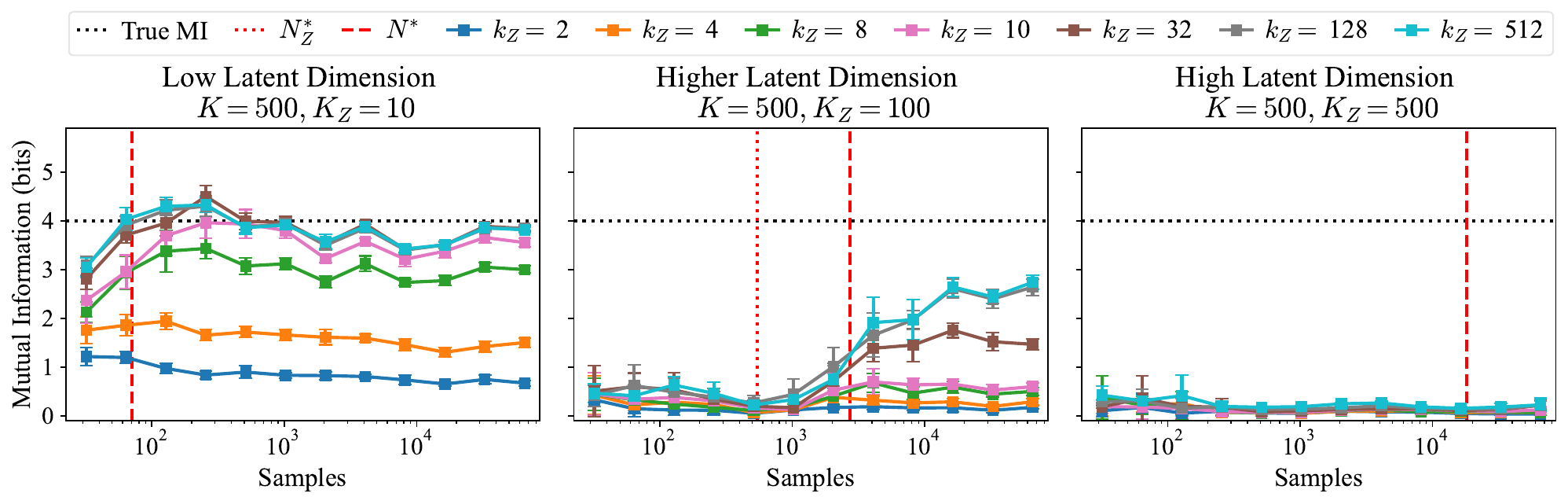}
\caption{\textbf{Effect of latent and critic dimensionality on InfoNCE.}  Curves show mean $\pm$ s.d. over ten runs. Panel represents $K=500$-dimensional data generated by teacher networks with latent dimensionality $K_Z = 10, 100,$ and $500$ (left to right). The true MI is 4 bits throughout. A sufficiently expressive critic ($k_Z \ge K_Z$) is required to recover all the information, yet the estimate approaches 4 bits only in the low-dimensional latent case ($K_Z \ll K$) when sample size satisfies $N \gg K_Z$. For larger latent spaces, the estimate remains far below the target even with large $N$. Vertical lines mark  $N$ needed for detection of nonzero MI using Gaussian random matrix models in the latent space ($N^*_Z$) and in the full space ($N^*$), cf.~Appx.~\ref{sec:SpikeDetection} ($N^*_Z\approx 1$ not shown in the left panel; and $N^*_Z=N^*$ in the right panel). Since a nonzero estimate emerges at $N>N^*_Z$, but $N<N^*$ if $K_Z\ll K$,  sampling of the latent space (not the full data space) governs the estimation even in the non-Gaussian setting.}
\label{fig:high_dim_finite_summarizable}
\end{figure}

Figure~\ref{fig:high_dim_finite_summarizable} shows results for InfoNCE, but other estimators behave similarly. Three regimes can be clearly seen. First, when the latent space is small-dimensional ($K_Z=10$, left), a sufficiently expressive critic with $k_Z \ge K_Z$ captures all dependencies once $N \gg K_Z$, and the estimate converges to the truth well below the sample threshold for detecting a spike in the full ambient space. In other words, the latent and not the ambient space dimensionality again governs when reliable estimation becomes possible. In the second regime with a moderate latent dimension ($K_Z=100$, center), $k_Z < K_Z$ fails outright, while $k_Z \ge K_Z$ retrieves only part of the information even at $N \sim 10^5$. This is because the signal per latent dimension is weaker ($\rho \approx 0.23$ vs.\ $0.65$), making each component harder to detect. Finally when the latent and observed dimensions coincide ($K_Z = K = 500$, right), neither an oversized critic nor $N \sim 10^5$ samples suffices for reliable estimation. Here, each canonical pair carries only $\sim 0.008$ bits, far below the detection threshold at any accessible $N$. And there is no longer the latent-space advantage because the data has no low-dimensional structure to exploit. The relationship between MI and embedding dimension also enables dimensionality estimation: the value of $k_Z$ at which MI saturates identifies the intrinsic latent dimensionality of the data \citep{gulati2026mutual}.

The transition between these regimes is relatively sharp. Below $N^*_Z$, estimates are near zero regardless of $k_Z$. Above $N^*_Z$, estimates grow and eventually plateau near the truth (for small $K_Z$) or continue rising slowly (for moderate $K_Z$). This rapid transition behavior can be argued for using random matrix theory method for linear models (Proposition~\ref{prop:sample_complexity}, Appx.~\ref{sec:SpikeDetection}). Surprisingly, it holds even in our nonlinear, non-Gaussian teacher network setting, suggesting the latent-space dimensionality based MI detection threshold may be a rather general phenomenon.

In summary, we see that, at least in our models, accurate MI estimation requires: (i)~statistical dependencies in the data distribution representable in a low-dimensional latent space; (ii)~a sufficiently expressive critic; and (iii)~enough data to sample that latent space, with the  scaling  $N \gtrsim K_Z^2/I$ (Proposition~\ref{prop:sample_complexity}). In other words, in high dimensions existence of accurate compressive embeddings is essential for reliable MI estimation from data sets of realistic sizes.

\begin{figure}[t]
    \centering
    \includegraphics[width=0.7\linewidth]{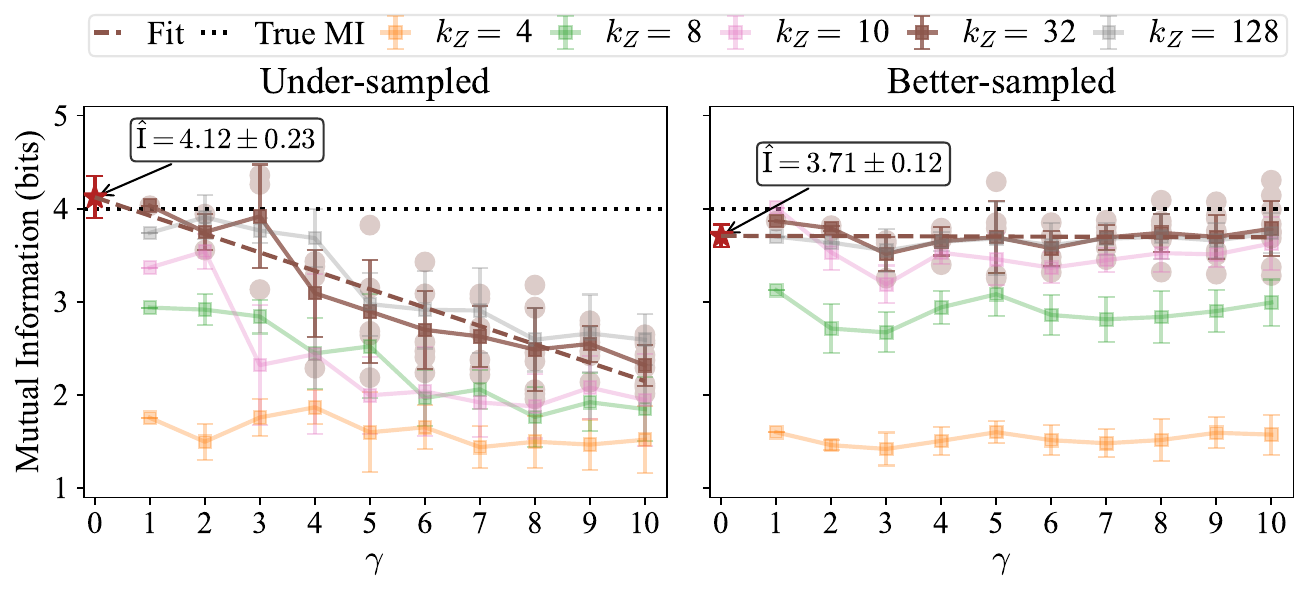}
    \caption{\textbf{Workflow for  MI estimation.} We apply InfoNCE with a separable critic to data from a random-teacher model ($K_X=K_Y=500$, $K_Z=10$, true MI 4 bits). (left) Undersampled, $N=256$. (right) Well-sampled, $N=2^{14}=16{,}384$. For each panel, MI is computed on $\gamma$ equal random, non-overlapping subsets (shown as circles for $k_Z=32$); error bars are $\pm$ s.d.\ over subsets. Estimates plateau at $k_Z = 32$ (used in final fits), indicating sufficient expressivity.  A linear fit to $\gamma\to0$ (infinite-data limit) gives the reported MI value.  The extrapolation removes sample-size dependent bias in the undersampled case; no such bias is visible in the well-sampled regime. }
    \label{fig:subsampling_sufficiency}
\end{figure}

\subsection{Practical Guide to MI Estimation} 
\label{txt:guide}
Because no estimator is unbiased for all distributions \citep{paninski2003estimation}, sample-size dependent bias is the dominant practical problem. The remedy is empirical: test how the estimate changes when the sample size  varies \citep{strong1998entropy,nemenman2004entropy,tang2014millisecond}. The {\em best practice} randomly partitions the data into $\gamma = 1,2,3,\dots$ equal subsets and computes  MI for each subset, $I_\mu(\gamma)$, $\mu = 1,\dots,\gamma$. One  checks if $I_\mu(\gamma)$ is statistically stable as $\gamma$ varies \citep{holmes2019estimation}, or if it extrapolates linearly to the hypothetical $\gamma = 0$ limit (effectively, infinite data) \citep{strong1998entropy}.  Large curvature in $I_\mu(\gamma)$ vs $\gamma$ signals that the estimator is not in its  asymptotic regime and is unreliable. In addition, the scatter of $I_\mu(\gamma)$ at fixed $\gamma$ provides an empirical variance that can be extrapolated to $\gamma = 1$ (all data) or $\gamma = 0$ (infinite data)  \citep{holmes2019estimation}.

We combine these best practices with our analysis into the following  guidelines for neural MI estimation (see Appx.~\ref{si_guideline} for details):
\begin{enumerate}
    \item Select an estimator suited to the expected MI range and data type. If the MI range is unknown, start with InfoNCE: it is stable and low-variance. If InfoNCE saturates near $\log(\text{batch size})$, the true MI likely exceeds that bound. Then switch to SMILE or SMILE--VSIB and increase the batch size. If InfoNCE and SMILE give substantially different estimates, the regime is likely high-MI or high-variance, and SMILE--VSIB is then the recommended choice. 
    \item Choose a critic network architecture that matches the data: a multilayer perceptron for generic data, a convolutional network for images, a transformer for sequences, etc.
    \item For $\gamma = [1,10]$ compute $I^{k_Z}_{{\rm EST},\mu}(\gamma)$ starting with $k_Z=1$ using the max-test early-stopping.
    \item Estimate $\bar{I}^{k_Z}_{{\rm EST}}(\gamma)\pm \sigma^{k_Z}(\gamma)$ as sample mean and standard deviation over subsets at fixed $\gamma$. 
    \item Increase $k_Z$ and repeat steps 3–4 until $\bar I^{k_Z}_{\rm EST}(\gamma)$ vs $k_Z$ no longer rises significantly.  The smallest dimension that reaches this plateau is $k^*_Z$; modestly over-estimating $k^*_Z$ is safe.
    \item If $I^{k^*_Z}_{{\rm EST},\mu}(\gamma)$ vs  $\gamma$ is approximately linear, extrapolate  to $\gamma\to0$ (details in Appx.~\ref{si_fit}).
    \item Report the extrapolated value $\hat{I}$ as the MI estimate together with its prediction interval $\Delta I$. If linear extrapolation is impossible, report failure to estimate. 
    \end{enumerate}

Figure~\ref{fig:subsampling_sufficiency} illustrates the workflow on data from a random teacher model with $K=500$, $K_Z=10$ and true MI  of 4 bits. The left and the right panels show the undersampled ($N=256$) and the well-sampled ($N=2^{14}$) cases, respectively. On the left, using $k^*_Z=32$, we can reliably extrapolate $I_{\text{InfoNCE}}$ to  $N\to\infty$ limit ($\gamma=0$) via subsampling. On the right, the estimator is already stable at this $N$; its slight downward bias matches Fig.~\ref{fig:high_dim_finite_summarizable} (see Discussion). We stress how striking this result is: in a 500$\times$500-dimensional, highly nonlinear setting (but with a 10-d latent structure), we obtain a near-perfect MI estimate, complete with accurate error bars, from just 256 samples! 

\subsection{Evaluations}
\paragraph*{Evaluations I: Standard Benchmarks.}
We evaluate our pipeline on the full 40-dataset benchmark suite introduced by \citet{czyz2024beyond}, which has become the standard reference for comparing MI estimators. These datasets span dimensions $1 \leq K \leq 50$ and include diverse dependence structures: Gaussian, Student-$t$, uniform, half-cube, spiral, and Swiss roll transforms. Using the same concatenated critic architecture as \citet{czyz2024beyond} (two hidden layers, width 16), but applying our bias correction and max-test stopping pipeline (9k train, 1k test samples), our method consistently matches or exceeds baseline InfoNCE accuracy while providing confidence intervals and internal consistency checks. Full results are in Appx.~\ref{appx:benchmark} (Table~\ref{tab:mi_results}), and a representative subset is shown in Table~\ref{tab:mi_results_best}.

For the most challenging tasks---spirals with nonlinear transformations---a more expressive critic (width 256) substantially improves performance, with our protocol correctly identifying when the smaller critic is insufficient
(flagged via the $\delta$ diagnostic). This illustrates the protocol's self-diagnostic value: it not only estimates MI but signals when the chosen critic architecture is inadequate (Appx.~Table~\ref{tab:mi_results_improved}).

\begin{table}[t]
\centering
\caption{\textbf{Representative results from the \citet{czyz2024beyond} benchmark suite} (results in \emph{nats} to match the reference). Our protocol consistently matches or exceeds simple InfoNCE while uniquely providing confidence intervals. Full results across all 40 tasks are in Appx.~Table~\ref{tab:mi_results}.}
\label{tab:mi_results_best}
\small
\setlength{\tabcolsep}{8pt}
\renewcommand{\arraystretch}{1.1}
\begin{tabular}{p{5.5cm}|
                >{\centering\arraybackslash}m{0.6cm}
                >{\centering\arraybackslash}m{0.8cm}
                >{\centering\arraybackslash}m{2.0cm}}
\toprule
Task & \makecell{True\\MI} & \makecell{Simple\\InfoNCE} & Ours \\
\midrule
Asinh @ Student-$t$ $3{\times}3$  & 0.29 & 0.20 & \textbf{0.26 $\pm$ 0.03} \\
Multinormal $25{\times}25$        & 1.29 & 1.20 & \textbf{1.26 $\pm$ 0.03} \\
Multinormal $50{\times}50$        & 1.62 & 1.40 & \textbf{1.60 $\pm$ 0.03} \\
Normal CDF @ Multinormal $25{\times}25$ & 1.02 & 0.80 & \textbf{0.94 $\pm$ 0.12} \\
Spiral @ Multinormal $25{\times}25$     & 1.02 & 0.70 & \textbf{0.97 $\pm$ 0.24} \\
Student-$t$ $3{\times}3$ (dof=2) & 0.29 & 0.20 & \textbf{0.26 $\pm$ 0.04} \\
Student-$t$ $3{\times}3$ (dof=3) & 0.18 & 0.10 & \textbf{0.18 $\pm$ 0.03} \\
\bottomrule
\end{tabular}
\end{table}

\paragraph*{Evaluations II: A Real-World Example, Noisy MNIST.}
We apply our pipeline to a realistic  {\em noisy MNIST} dataset \citep{Haffner1998, Bilmes2015, Livescu2016, abdelaleem2025deep}. Each sample consists of two $28\times28=784$ dimensional views: $X$ is a randomly rotated and scaled image of a digit, and $Y$ is another random digit with the same label overlaid with Perlin noise.  The only shared information is the digit label (10 classes), giving the ground truth MI of $\log_{2}10 \approx 3.3$ bits.  The data dimension is  $K=784$, while the latent dimension is low (but unknown) because only ten classes matter. Random matrix analysis, Appx.~\ref{sec:SpikeDetection}, suggests that the nonzero MI detection thresholds are $N^*_Z\approx17$ (for $K_Z=10)$ or $N^*\approx 2700$. Numerical experiments show that $N\sim512$ is sufficient for detection, so that, again, good sampling in the latent (not data) space is sufficient. Here, however, we need {\em accurate} estimation of MI, rather than just significant difference from zero.

\begin{figure}[t]
    \centering
    \includegraphics[width=0.4\textwidth]{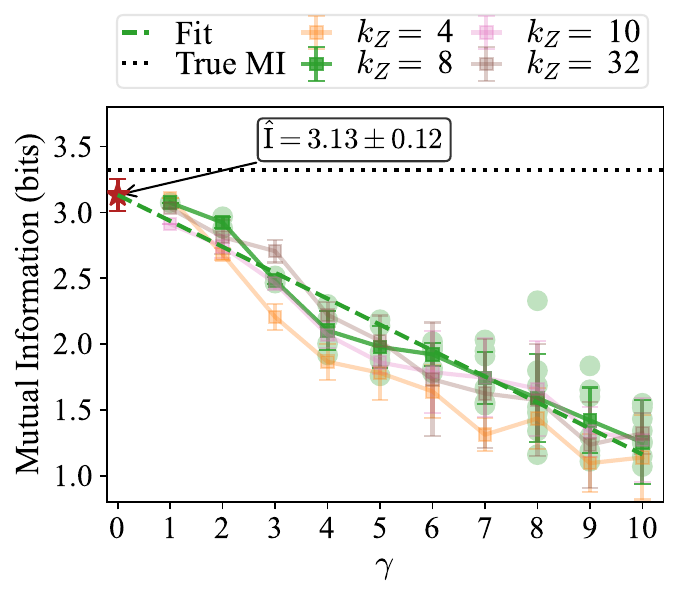}
    \caption{\textbf{MI estimation on noisy MNIST.} We apply InfoNCE with a separable critic to $N=2^{14}=16{,}384$ samples from the noisy MNIST dataset ($K_X=K_Y=784$, true MI $\approx 3.3$ bits). MI is computed on $\gamma$ equal non-overlapping subsets for increasing embedding dimensions $k_Z$. Estimates plateau at $k_Z=32$, indicating sufficient expressivity. Linear extrapolation to $\gamma\to0$ yields $\hat{I}=3.13\pm0.12$ bits, within two error bars of the ground truth. A slight kink at $\gamma=4$ indicates mild underestimation for $N\lesssim 4\times10^3$, likely attributable to the clustered, non-smooth latent structure of digit classes.}
    \label{fig:mnist_case_study}
\end{figure}

With $N = 2^{14}=16{,}384$ samples, reliable linear extrapolation to $\gamma=0$ yields $\hat{I} =3.13\pm 0.12$ bits, within two error bars of the true MI (Fig.~\ref{fig:mnist_case_study}). A slight kink at $\gamma=4$ indicates that the estimator will underestimate for $N\lesssim 4\times10^{3}$, likely attributable to the clustered, non-smooth latent structure of digit classes rather than the ambient dimension. Reliable estimation on noisy MNIST thus
requires several thousand examples---not the hundreds of thousands that traditional methods would need in 784 dimensions!

\paragraph*{Evaluations III: Scaling to Natural Images, CIFAR-10/100.}
We further validate the protocol on noisy versions of CIFAR-10 and CIFAR-100 \citep{krizhevsky2009learning}, where $K=3072$ ($32\times32$ RGB images). As with noisy MNIST, each sample consists of two views of the same class with independent noise applied, so the only shared information is class identity. The theoretical upper bound on MI is $\log_2 C$ bits, where $C$ is the number of classes, assuming classes are perfectly distinct. This  assumption holds less cleanly for CIFAR than for MNIST due to  similarity across some classes.

\begin{figure}[t]
    \centering
    \includegraphics[width=0.7\textwidth]{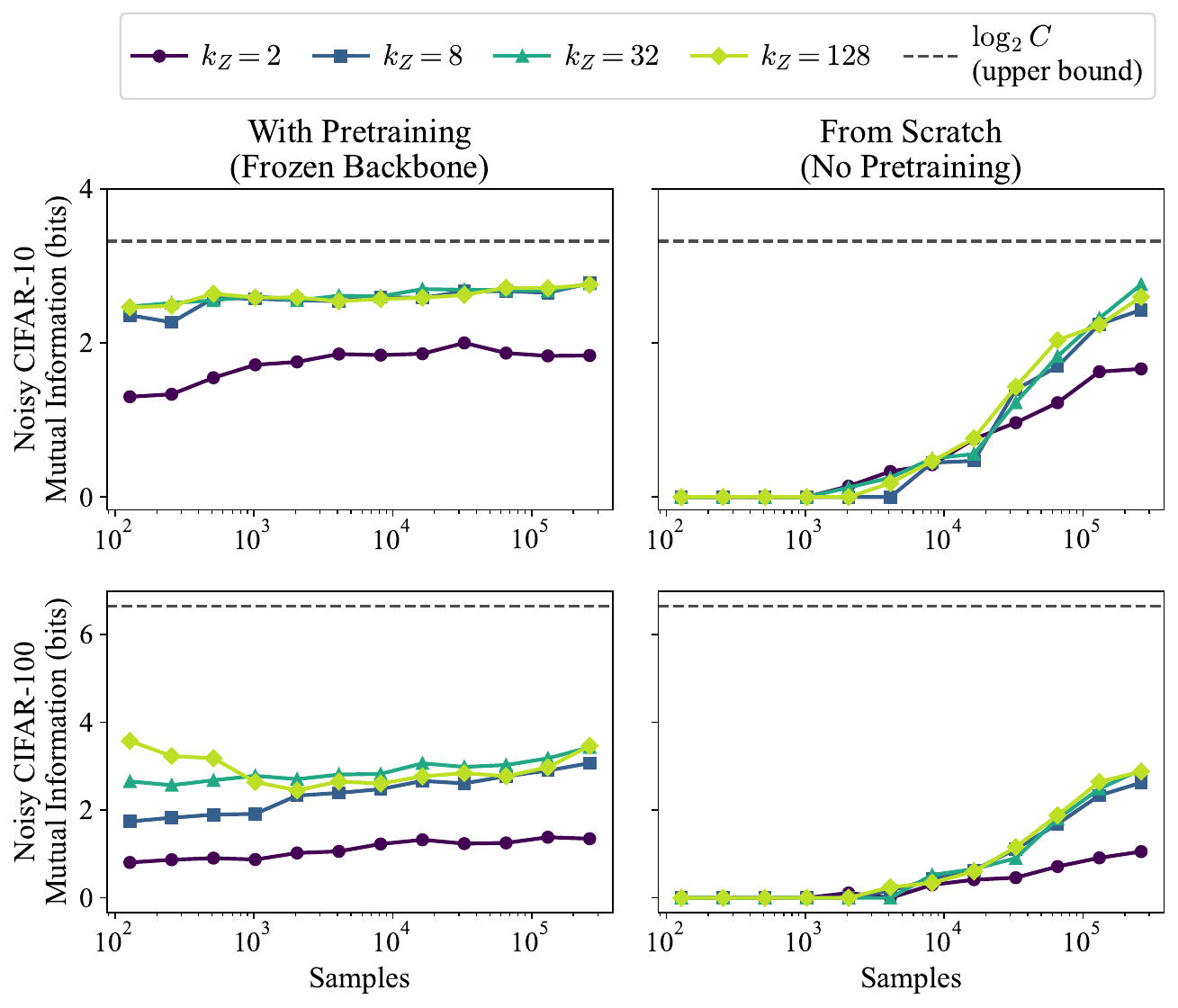}
    \caption{\textbf{MI estimation on noisy CIFAR-10 and CIFAR-100.} MI is estimated as a function of sample size $N$ for different critic embedding dimensions $k_Z$, under two conditions: a frozen pretrained ResNet-20 backbone (left column) and end-to-end training from scratch (right column). Top row: CIFAR-10 ($C=10$ classes); bottom row: CIFAR-100 ($C=100$ classes). Dashed horizontal lines show $\log_2 C$ bits, the theoretical upper bound assuming perfectly distinct classes. With a frozen backbone, MI stabilizes rapidly and becomes independent of $N$, confirming that the backbone has already resolved the relevant latent structure. From-scratch estimates grow with $N$ but approach the same saturation values, consistent with the frozen backbone serving as an empirical ceiling. CIFAR-100 saturates particularly far below $\log_2 100$ bits due to similarity across the data classes. The ResNet-20 backbone uses pretrained weights from \citet{chenyaofo2021cifar}.}
    \label{fig:cifar_results}
\end{figure}

Figure~\ref{fig:cifar_results} shows MI estimates as a function of sample size $N$ for different critic embedding dimensions $k_Z$, under two conditions: training the critic on top of a frozen pretrained
ResNet-20 backbone \citep{he2016deep}, and training the full pipeline (ResNet-20 and the critic) from scratch. In the from-scratch setting, MI estimates grow with $N$ and do not plateau even by $N\sim10^5$, consistent with the larger sample requirements expected for natural image data relative to MNIST. Crucially, the frozen backbone condition reveals that these estimates are approaching a genuine ceiling: with the backbone fixed and only the projection head trained (cf. Appx.~\ref{app:cifar}), MI stabilizes quickly and becomes essentially independent of $N$, so that the backbone pre-trained on far more data has already compressed the images into a latent representation that the critic can then resolve with many fewer samples. The saturation value for the frozen backbone ($\sim$3 bits for CIFAR-10, $\sim$4 bits for CIFAR-100) thus serves as an empirical upper bound on the achievable MI given the noise structure and class similarity of these datasets.  That CIFAR-100 saturates well below $\log_2 100 \approx 6.6$ bits particularly reflects the substantial similarity across its 100 fine-grained classes, which reduces the true shared information. This suggests that the from-scratch curves would eventually plateau near these values with sufficient data. Overall, these results confirm that sample complexity being dictated by just  the latent-space dimension extends to genuinely complex natural image data and deep convolutional backbones. Further, they show that prior knowledge encoded in a pretrained model can dramatically reduce the samples needed for reliable MI estimation.

\section{Discussion}
\label{sec:discussion}

Accurate MI estimation for high-dimensional data remains difficult because no estimator is uniformly unbiased \cite{paninski2003estimation}.  Neural estimators offer promise, but most studies test them in well-sampled, linear settings that hide their limitations.  We show that neural MI estimators become reliable when the underlying statistical dependence  between the variables lies in a low-dimensional latent space, the chosen critic is expressive enough to capture that dependence, and the dataset is large enough to resolve it in the latent space.  We propose three improvements that make neural MI estimation practical under these conditions.  First, a ``max-test’’ early-stopping rule prevents the runaway growth common in estimators based on DV objectives.  Second, using the MINE family critics within a probabilistic VSIB wrapper regularizes the estimation and decreases fluctuations at higher information values (up to $\sim10$ bits in our experiments).  Third, a subsampling-and-extrapolation workflow can be used to detect bias, choose the right critic dimension, and estimate error bars. Although we illustrated this with InfoNCE and specific network architectures, our analysis  can incorporate any DV-based estimator.

Using these ingredients, we obtained near-exact MI estimates, with error bars, in regimes that defeat most methods (Appx.~Fig.~\ref{si_fig:KSG_examples}), e.g., 784-dimensional images with only $\sim 10^4$ samples. Across all experiments, our procedure never significantly overshot the ground truth MI values, which is crucial to avoid false positives in scientific applications \citep{strong1998entropy,margolin2006aracne,tang2014millisecond}. In contrast, modest underestimation vanishes as $N$ grows and finer structure is learned, and is, therefore, not a major concern.

\paragraph*{Limitations.}
The protocol's reliability rests on the low-dimensional latent structure assumption: if the true $K_Z$ is comparable to $K$, sample complexity is too high and estimation may be infeasible at any accessible $N$ (Fig.~\ref{fig:high_dim_finite_summarizable}, right panel). In such cases, the protocol correctly signals failure via the nonlinear $\gamma$ curve, but provides no reliable estimate. Second, all theoretical analysis is done under a Gaussian latent variable model. While our empirical evidence suggests the qualitative conclusions transfer to nonlinear settings, formal guarantees in the non-Gaussian case remain an open problem. Third, our validation uses settings where the ground truth MI is known or tightly bounded. In contrast, applying the protocol to real scientific data requires careful choices about data representation that are domain-specific and beyond the scope of this work. Some of these applications are the subject of ongoing and published follow-on research \citep{gulati2026mutual}. Finally, the stopping heuristic and the subsampling extrapolation procedure are justified under the assumption that the trained critic is a sufficiently good approximation of the globally optimal critic. When this assumption fails (e.~g., when the critic class is too restricted or training is unstable), the reported estimate may still be unreliable even if the $\gamma$ curve appears linear.

High-dimensional MI estimation will always require careful diagnostics and a measure of skepticism. And yet, our approach turns neural MI estimators into practical tools for regimes that were previously inaccessible, $N \lesssim K$, with the potential to substantially broaden how MI estimation is used in scientific research.

\acks{We thank Sean Ridout, Michael Pasek, Paarth Gulati, Stephanie Palmer, and Gasper Tkacik for useful discussions. This work was supported, in part, by NSF grants 2010524 and 2409416, by NIH grant NS084844, and by the Simons Foundation Investigator program.}
\newpage

\bibliography{refs}
\newpage
\pagenumbering{roman} 
\setcounter{page}{1}
\renewcommand{\theequation}{\thesection.\arabic{equation}}
\setcounter{equation}{0}
\appendix
\section*{Appendix Table of Contents}
\startcontents[appendices]
\printcontents[appendices]{}{0}{\setcounter{tocdepth}{2}}

\section{Theoretical Background}
\label{app:theory}
\subsection{Deriving the DV-Based Estimators (MINE, SMILE, InfoNCE)}
\label{app:dv_derivations}

Two widely used neural estimators of mutual information—InfoNCE and SMILE—can be derived from the Donsker–Varadhan (DV) representation \citep{donsker1983asymptotic} of the Kullback–Leibler (KL) divergence.

Mutual information (MI) between random variables $X$ and $Y$ can be expressed as the KL divergence between their joint distribution and the product of marginals:
\begin{equation}
    I(X; Y) = D_{\mathrm{KL}}(p(x, y) \| p(x)p(y)) = \mathbb{E}_{p(x, y)} \left[ \log \frac{p(x, y)}{p(x)p(y)} \right].
    \label{eq:mi_joint}
\end{equation}
Alternatively, MI can be factorized using the conditional distribution:
\begin{equation}
    I(X; Y) = \mathbb{E}_{p(x)} \left[ D_{\mathrm{KL}}(p(y|x) \| p(y)) \right]
    = \mathbb{E}_{p(x)} \left[ \mathbb{E}_{p(y|x)} \left[ \log \frac{p(y|x)}{p(y)} \right] \right].
    \label{eq:mi_conditional}
\end{equation}

The DV representation provides a lower bound on KL divergence between two distributions $P$ and $Q$:
\begin{equation}
    D_{\mathrm{KL}}(P \| Q) \geq \sup_T \left[ \mathbb{E}_P[T] - \log \mathbb{E}_Q[e^T] \right],
\end{equation}
where the supremum is over all measurable functions $T$ such that $\mathbb{E}_Q[e^T] < \infty$. Equality is achieved when $T^* = \log \frac{dP}{dQ} + c$ for any constant $c$.

Applying the DV representation to Eq.~\eqref{eq:mi_joint}, one obtains:
\begin{equation}
    I(X;Y) \geq I_{\rm MINE}(X; Y) \coloneq \max_T \left[ \mathbb{E}_{p(x,y)}[T(x, y)] - \log \mathbb{E}_{p(x)p(y)}\left[ e^{T(x, y)} \right] \right],
    \label{eq:mine2}
\end{equation}
where $T(x, y)$ is a learned critic function approximating the log-density ratio. This is the MINE estimator \citep{Belghazi2018MutualEstimation, poole2019variational}, which exhibits large variance empirically, specifically because of the second term. 

To stabilize this estimator, SMILE \citep{song2019understanding} clips the critic before exponentiation to reduce the influence of outliers on the normalization term:
\begin{equation}
    I(X;Y)\ge I_{\mathrm{SMILE}}(X; Y) \coloneq \max_T \left[ \mathbb{E}_{p(x,y)}[T(x, y)] 
    - \log \mathbb{E}_{p(x)p(y)}\left[ e^{\text{clip}(T(x, y), -\tau, \tau)} \right]\right],
    \label{eq:smile2}
\end{equation}
where $\text{clip}(z, -\tau, \tau) = \min(\max(z, -\tau), \tau)$, and $\tau > 0$ effectively controls the bias–variance trade-off. 

Using the conditional factorization of MI in Eq.~\eqref{eq:mi_conditional}, one can again apply the DV representation, now to $D_{\mathrm{KL}}(p(y|x) \| p(y))$:
\begin{equation}
    D_{\mathrm{KL}}(p(y|x) \| p(y)) \geq \sup_{T_x} \left[ \mathbb{E}_{p(y|x)}[T(x, y)] 
    - \log \mathbb{E}_{p(y)}[e^{T(x, y)}] \right].
    \label{eq:dv_conditional}
\end{equation}
Plugging into Eq.~\eqref{eq:mi_conditional} and swapping the order of the integrals gives:
\begin{equation}
    I(X; Y) \geq \sup_T \left[ \mathbb{E}_{p(x, y)}[T(x, y)] 
    - \mathbb{E}_{p(x)} \left[ \log \mathbb{E}_{p(y)} \left[ e^{T(x, y)} \right] \right] \right],
    \label{eq:infonce_bound}
\end{equation}
which is the first step towards the InfoNCE estimator \citep{oord2018representation}.

In practice, the expectation over $p(y)$ is approximated using contrastive sampling. For a batch $\{(x_i, y_i)\}_{i=1}^N$:
\begin{itemize}[noitemsep]
    \item Treat $y_i$ as a \emph{positive} sample from $p(y|x_i)$,
    \item Treat $\{y_j\}_{j \neq i}$ as \emph{negative} samples from $p(y)$.
\end{itemize}
Using a Monte Carlo approximation of the expectations, one obtains:
\begin{equation}
    I(X;Y)\ge I_{\mathrm{InfoNCE}}(X; Y) \coloneq 
     \frac{1}{N} \sum_{i=1}^N 
    \log \frac{e^{T(x_i, y_i)}}{\frac{1}{N}\sum_{j=1}^N e^{T(x_i, y_j)}}.
    \label{eq:infonce2}
\end{equation}

\subsection{Bilinear Critics, Gaussian Variables, and \texorpdfstring{$I_{\rm CCA}$}{ICCA}}
\label{app:cca_derivation}

It is well known that, for jointly Gaussian variables, MI between two random variables can be written in terms of their nonzero canonical correlations, subject to keeping enough canonical pairs \citep{kullback1959information, gelfand1959calculation, huffmann2022distribution}:
\begin{equation}
    I_{\rm CCA}=-\frac{1}{2}\sum_{i}^{K_Z}\log(1-\rho_i^2),\label{eq:I_CCA}
\end{equation}
where $\rho_i$ are the canonical correlations. Here, we show that this CCA estimate also emerges naturally as a special case of the \emph{DV bound} on MI, with the concatenated bilinear (quadratic) critic family.  This connects CCA to the same variational estimator framework as neural methods like InfoNCE and MINE, but with a different critic class.

Let \( X \in \mathbb{R}^{K_X} \) and \( Y \in \mathbb{R}^{K_Y} \) be jointly Gaussian with:
\begin{equation}
\begin{bmatrix} x \\ y \end{bmatrix} \sim \mathcal{N}\left(0, \begin{bmatrix} \Sigma_{XX} & \Sigma_{XY} \\ \Sigma_{YX} & \Sigma_{YY} \end{bmatrix}\right)\coloneq \mathcal{N}\left(0,\Sigma\right).
\end{equation}
Using the same DV representation as for the MINE estimator, Eq.~(\ref{eq:mine2}), we write:
\begin{equation}
I(X;Y) \geq \max_{T} \left\{\mathbb{E}_{p(x,y)}[T(x,y)] - \log \mathbb{E}_{p(x)p(y)}[e^{T(x,y)}]\right\}.
\label{eq:DV_CCA}
\end{equation}
As always, the globally optimal critic saturating the bound is $T^*(x,y)=\log\frac{p(x,y)}{p(x)p(y)}+c$, where $c$ is an arbitrary constant. For Gaussian data, this expression becomes:
\begin{equation}
T^{*}(x,y)=\frac{1}{2}\left[
x^{\!\top}\Sigma_{XX}^{-1}x
+y^{\!\top}\Sigma_{YY}^{-1}y
-\begin{pmatrix}x\\y\end{pmatrix}^{\!\top}
\Sigma^{-1}
\begin{pmatrix}x\\y\end{pmatrix}\right]+c.
\label{eq:Topt1}
\end{equation}

We now define the canonical pairs. First, the whitened cross-covariance is 
\begin{equation}
{\mathcal K}=\Sigma_{XX}^{-1/2}\Sigma_{XY}\Sigma_{YY}^{-1/2}
=U\Lambda V^{\top},\qquad
\Lambda=\operatorname{diag}(\rho_{1},\ldots,\rho_{K_Z}),
\end{equation}
where $U$ and $V$ are the matrices of left and right singular vectors of ${\mathcal K}$, and  $\Lambda$ has canonical correlations on the diagonal. Then the canonical coordinates are
\begin{equation}
    u=U^\top \Sigma_{XX}^{-1/2}x,\quad     v=V^\top \Sigma_{YY}^{-1/2}y.
\end{equation}
 
In these coordinates, the optimal critic is just the sum over independent canonical pairs:
\begin{equation}
T^{\star}(x,y)\;=\;
\sum_{i=1}^{K_Z}\frac{\rho_{i}}{1-\rho_{i}^{2}}
\left[u_{i}v_{i}-\frac{\rho_{i}}{2}\big(u_{i}^{2}+v_{i}^{2}\big)\right]
\label{eq:Topt2}
\end{equation}
(the expression under the sign of the sum can be verified by direct calculation for a bivariate normal distribution over $u_i$, $v_i$). Plugging Eq.~(\ref{eq:Topt2}) into Eq.~(\ref{eq:DV_CCA}) then gives MI in the form Eq.~(\ref{eq:I_CCA}).

In other words, the DV optimal critic for the Gaussian distribution is a quadratic form, Eq.~(\ref{eq:Topt2}). Such a form belongs to a class of bilinear concatenated critics in Eq.~(\ref{eq:generalized_critic}), where $f$ concatenates its arguments and forms a bilinear expression from them, and $g$ and $h$ are linear operators, or simply identities. 

We can also achieve the same result by directly calculating the optimal critic within a family of concatenated quadratic forms
\begin{equation}
    T= z^\top W z,\quad z^\top=(x^\top,y^\top),
\end{equation}
with the matrix $W$ such that both terms in the r.h.s.\ of Eq.~(\ref{eq:DV_CCA}) are finite. Then
\begin{align}
\mathbb E_{p(x,y)}[z^{\top}Wz]&=\operatorname{tr}(\Sigma W),\\
\mathbb E_{p(x)p(y)}e^{z^{\top}Wz}
&=\frac{1}{\sqrt{\det\left(I-2\Sigma_{\text{prod}}W\right)}},
\quad \text{where }
\Sigma_\text{prod}= \begin{bmatrix} \Sigma_{XX} & 0 \\ 0& \Sigma_{YY} \end{bmatrix},
\end{align}
provided  $I-2\Sigma_\text{prod}W$ is positive definite. Now differentiating the DV bound, Eq.~(\ref{eq:DV_CCA}) w.r.t.\ $W$, we find the condition for $W^\star$, which optimizes the critic: 
\begin{equation}
\Sigma-\left(\Sigma_{\text{prod}}^{-1}-2W^\star \right)^{-1}=0.
\end{equation}

This results in 
\begin{equation}
W^{\star}=\frac{1}{2}\left(\Sigma_{\text{prod}}^{-1}- \Sigma^{-1}\right),
\end{equation}
which, sandwiched between $z^\top$ and $z$, again gives Eqs.~(\ref{eq:Topt1}) and~(\ref{eq:Topt2}).

Overall, these results say that the CCA estimate of MI  emerges naturally from the bilinear (quadratic) concatenated critic family for Gaussian data within the DV framework.

\subsection{Probabilistic Critics: Variational Symmetric Information Bottleneck (VSIB)}
\label{app:vsib}

The Variational Symmetric Information Bottleneck (VSIB)\footnote{This is an instance of the more general, Deep Multivariate Information Bottleneck Framework \citep{Tishby2013, abdelaleem2025deep}. In this framework, one specifies a compression/encoder graph that is traded off against a generative/decoder graph. Each graph is then transformed into an information bound that can be optimized.} can formalize MI estimation as a form of probabilistic dimensionality reduction. It introduces a latent representation for each variable—$Z_X$ and $Z_Y$—produced by separate stochastic encoders from $X$ and $Y$, respectively. Mutual information is then estimated between these latent representations using {\em any} neural MI estimator of choice.

This leads to the following objective \citep{abdelaleem2025deep}:
\begin{equation}
\label{eq:vsib2}
L_{\text{EST}_\text{VSIB}} = I^E(X; Z_X) + I^E(Y; Z_Y) - \beta I^D_{\text{EST}}(Z_X; Z_Y),
\end{equation}
where $I^E(\cdot\,;\cdot)$ are encoder regularization terms, and $I^D_{\text{EST}}(Z_X; Z_Y)$ is the mutual information in the latent space estimated using a particular chosen neural estimator (e.g., InfoNCE or SMILE).

Each encoder term is computed as:
\begin{align}
I^{E}(X;Z_X) &\approx\frac{1}{N}\sum_{i=1}^N D_{\rm KL}(p(z_x|x_i) \Vert r(z_x)) \nonumber \\&\approx \frac{1}{2N}\sum_{i=1}^N \left[\text{Tr}({\Sigma_{Z_X}(x_i)}) +||\vec{\mu}_{Z_X}(x_i)||^2-k_{Z_X}-\ln \det(\Sigma_{Z_X}(x_i)) \right],\label{IExzx} 
\end{align}
where $\vec{\mu}_{Z_X}(x)$ and $\Sigma_{Z_X}(x)$ parameterize the mean and covariance of the encoder distribution $p(z_x|x)$, and $k_{Z_X}$ is the latent dimensionality. The same form is used for $I^E(Y; Z_Y)$. The scalar $\beta > 0$ controls the trade-off between the regularization terms and the estimated information. In our experiments, we used $\beta = 512$, which strongly prioritizes $I^D_{\text{EST}}$ while still regularizing the encoder mappings. Other large values of $\beta$ yielded similar results (not shown).

In the limit where the encoders $p(z_x|x)$ and $p(z_y|y)$ collapse to delta distributions, the $I^E$ terms converge to the entropy $H(z_x)$ and $H(z_y)$, respectively. These terms diverge to infinity and do not change as a function of the embedding and thus do not affect the estimation of information between the latent variables, so that Eq.~(\ref{eq:vsib2}) recovers the standard deterministic neural MI estimators.

While VSIB naturally aligns with separable critics (i.e., $T(x,y) = \langle g(x), h(y)\rangle$), the framework can be generalized to concatenated critics. Specifically, one can define a latent variable $Z$ such that:
\begin{equation}
Z \sim \mathcal{N}(\mu([x, y]), \Sigma([x, y])),
\end{equation}
with neural networks $\mu(\cdot)$ and $\Sigma(\cdot)$ operating on the joint input $[x, y]$. The loss becomes:
\begin{equation}
L = I^E([X,Y]; Z) - \beta I^D_{\text{EST}}(Z(X,Y)),
\end{equation}
where $I^D_{\text{EST}}(Z(X,Y))$ is just another way to write $I(X;Y)$ for a concatenated critic. While this version is less directly interpretable in terms of a variable compression, it is straightforward to implement and was used in Figs.~\ref{fig:low_dim_infinite_main} and \ref{fig:high_dim_infinite_main} as the probabilistic variant of SMILE and InfoNCE (denoted with $\text{EST}_\text{VSIB}$).

\subsection{Estimating Sample Size Required for MI Estimation in the Latent Variable Model}
\label{sec:SpikeDetection}

To make sense of the sample sizes in Fig.~\ref{fig:high_dim_finite_summarizable}, we connect our latent variable model to the spiked covariance model from random matrix theory~\citep{potters2020first}. In that model, the data covariance has structure $\Sigma = \sigma_n^2(\mathbb{I} + \theta v^\top v)$, where $\sigma_n^2$ is per-coordinate noise variance, $v$ is a low-rank ``spike'' matrix with orthonormal columns (the signal directions), and $\theta$ is the signal magnitude. The spike separates from sampling-induced background correlations when $\theta > \sqrt{k/N}$, where $k$ is the observed dimensionality \citep{baik2005phase}.

Our latent model can be cast in this language by considering $2K_Z$-dimensional Gaussian variables $(x^\top, y^\top)$, where each pair $(x_i, y_i)$ is produced from a latent variable $z_i$, with $\mathbb{E}x_ix_j = \mathbb{E}x_iy_j = \mathbb{E}y_iy_j = 0$ for $i \neq j$, and $v^\top = (1/\sqrt{2},\, 1/\sqrt{2})$. The covariance of the $i$th pair is
\begin{equation}
    \Sigma_i = \begin{pmatrix}
    1+\theta^\prime_i & \theta^\prime_i \\
    \theta^\prime_i   & 1+\theta^\prime_i
    \end{pmatrix},
\label{eq:singlesigma}
\end{equation}
with $\theta^\prime = \theta/2$ and correlation $\rho_i =
\theta^\prime_i/(1+\theta^\prime_i)$. Since $I_i = -\frac{1}{2}\log(1-\rho_i^2)$
for each pair,
\begin{equation}
\theta^\prime_i = \frac{1}{1 - (1-2^{-2I_i})^{1/2}} - 1.
\end{equation}

\begin{proposition}[Sample complexity in the latent variable model]
\label{prop:sample_complexity}
Consider the latent Gaussian model above with $K_Z$ correlated pairs, total MI $I$ distributed uniformly across pairs, and $K_Z \gg 1$, $\operatorname{rank} v \ll 2K_Z$. The minimum number of samples required for nonzero MI detection in the latent space is
\begin{equation}
N > N^*_Z = \frac{K_Z\!\left(1 - \sqrt{1 - 2^{-2I/K_Z}}\right)^2}
                 {2\!\left(1 - 2^{-2I/K_Z}\right)}.
\label{eq:NstarZ}
\end{equation}
In the weak-signal limit $2I/K_Z \ll 1$, this simplifies to
\begin{equation}
N \gtrsim \frac{K_Z^2}{I\ln 2}
          \!\left(1 - 2\sqrt{2\ln 2\, I/K_Z}\right),
\label{eq:NstarZ_weak}
\end{equation}
which scales as $N \sim K_Z^2/I$. Repeating the calculation in the full ambient space gives the looser bound $N > N^*$, corresponding to detecting a linear spike in the $K$-dimensional data space.
\end{proposition}

The quadratic scaling $N \sim K_Z^2$ arises because higher latent dimensionality both makes signal detection harder and reduces the information per dimension. Reliable MI estimation cannot begin until each spike is detected ($N > N^*_Z$), and $N$ must grow further beyond $N^*_Z$ for the detected eigenvalues to approach the true spike magnitudes, allowing MI to converge to its true value.

The bound $N^*_Z$ is optimistic: it ignores interactions among spikes and, for nonlinear variants of the model, also the cost of learning the nonlinear embedding from the $2K$-dimensional data space. Thus, one would naively expect $N \gg N^* > N^*_Z$ in practice. Fig.~\ref{fig:high_dim_finite_summarizable} shows the opposite: at least for the nonlinear teacher model, accurate MI becomes possible shortly after $N$ exceeds the tighter latent bound $N^*_Z$ and before $N^*$ is reached, at least for moderate $K_Z$. This is the central empirical finding of Sec.~\ref{txt:results}: sample complexity is governed by the statistical structure of the latent space, not the ambient space. Note that these conditions are not strictly satisfied in our experiments, so that the proposition provides a theoretical lower bound on sample complexity that is approximately predictive of empirical behavior.

\section{The Estimation Protocol: Justification and Details}
\label{app:protocol}

\subsection{Justification for the Max-Test Stopping Heuristic}
\label{app:explain_max_test_heuristic}
To define the stopping heuristic, note both the training and the test data are sampled from the same $p(x,y)$. In practice, expectations for every estimator EST, such as in Eqs.~(\ref{eq:mine2},~\ref{eq:smile2},~\ref{eq:infonce2}), are implemented with empirical sampling. That is, we form empirical densities $\pi_{\rm train} = \frac{1}{N_{\rm train}} \sum_{i=1}^{N_{\rm train}} \delta (x-x_i,y-y_i)$, and similar for $\pi_{\rm test}$ for the test data. Then, for example, the MINE estimator is implemented as, 
\begin{equation}
L_{\rm EST} (\pi_{\rm train},T)= \mathbb{E}_{\pi_{\rm train}(x,y)}[T(x,y)] - \log \mathbb{E}_{\pi_{\rm train}(x)\pi_{\rm train}(y)}\left[e^{T(x,y)}\right],\label{eq:Lmine}
\end{equation}
and similarly for the other estimators. Then the ``train'' and ``test'' MI values are defined via:
\begin{align}
T^*_{\rm train}&= \arg \max_T L_{\rm EST}(\pi_{\rm train},T),\\
I_{\text{EST, train}} &= L_{\rm EST} (\pi_{\rm train},T^*_{\rm train}),\\
I_{\text{EST, test|train}} &= L_{\rm EST} (\pi_{\rm test},T^*_{\rm train}).
\label{eq:mine_emp}
\end{align}
For completeness, we also define the true (typically unknown) mutual information as $I_{\rm true}$, with the globally optimal critic 
\begin{equation}
     T^*(x,y)=\log\frac{p(x,y)}{p(x)p(y)}+c,
\end{equation}
where $c$ is an arbitrary constant.

With these definitions, we use the following procedure:
\begin{enumerate}
\item Train the estimator on $\pi_{\rm train}$ for several epochs in each step of the algorithm, see Appx.~\ref{app:implementation}.
\item After each such step, freeze $T^*_{\rm train}$, and evaluate both $I_{\text{EST, train}}$ and $I_{\text{EST, test|train}}$\ for whichever estimator EST is being used. 
\item Select $\hat{T}=T^*_{\rm train}$ corresponding to the cycle where  $I_{\text{EST, test|train}}$ is maximal.
\item Report  $I_{\text{EST, train}}$ evaluated at $\hat{T}$ as the final MI estimate.
\end{enumerate}

This procedure regularizes overfitting. While one could attempt architectural regularization  to stabilize training (e.~g., dropout or weight decay), such strategies offer no clear way to assess the trustworthiness of the resulting estimate. In contrast, the max-test heuristic explicitly favors models that perform best on unseen data, and the shape of the test curve itself serves as a reliability diagnostic.

The procedure may appear to conflict with standard machine learning practice of reporting test-set results. The distinction arises because MI estimation is an \emph{estimation of a functional}, not a \emph{prediction} task. Because MI is a nonlinear functional of $p(x,y)$, unbiased estimates of the distribution do not yield unbiased MI estimates~\citep{nemenman2001entropy}, and resampling approaches such as bootstrap and cross-validation introduce systematic biases in MI estimation~\citep{holmes2019estimation}. The role of the test set in our heuristic is therefore not to provide the reported estimate, but to serve as a stopping signal, namely the epoch at which the test MI peaks identifies the best-generalizing critic resolution. We then  report the corresponding training MI. We justify this choice below, under the explicit assumption that the trained critic $T^*_{\rm train}$ is a sufficiently good approximation of the globally optimal critic $T^*$---an assumption that is empirically verifiable and supported by all experiments in this paper.

\paragraph*{Biases of estimators.}

General bounds showing that $I_{\rm EST, test|train}$ typically understimates $I_{\rm true}$ and hence should not be used as a reported estimate cannot exist without additional strong assumptions about $T$. To see this, note a counter-example: if the test set consists of just one sample, and the training set has many, and $T$ is optimized over the class that contains just a single peak, but at different locations in $(x,y)$, then the test MI can be very large (when the critic peak matches the single sample), while the training MI will be low, and either can over- or under-estimate $I_{\rm true}$. Thus to argue for the max-test heuristic, we need additional assumptions. 

There are many variants of similar such assumptions, all starting with assuming that  $T^*_{\rm train}\approx T^*$ (that is, the trained optimal critic is almost globally optimal), and all resulting in $I_{\rm EST, test|train}\le I_{\rm true}$. We do not know which of the assumptions would be convincing to the reader, and so here we give just one loose proof arguing that the test value is an underestimate and should not be reported.  Since test and training sets are taken from the same distribution, and assuming they are the same size
\begin{multline}
\mathbb{E}_{p(x,y)} I_{\rm EST, test|train} \equiv\mathbb{E}_{p(x,y)}L_{\rm EST} (\pi_{\rm test},T^*_{\rm train})\\ \le \mathbb{E}_{p(x,y)}L_{\rm EST} (\pi_{\rm test},T^*_{\rm test}) = \mathbb{E}_{p(x,y)}L_{\rm EST} (\pi_{\rm train},T^*_{\rm train}) \equiv \mathbb{E}_{p(x,y)}I_{\rm EST, train},
\end{multline}
where the inequality is due to the maximization in the definition of $I_{\rm EST}$. In other words, the test MI is expected to be lower than the training value for statistically similar test and training sets. In particular, this means that if $T^*_{\rm train}=T^*$ and $I_{\rm EST, train}=I_{\rm true}$, 
then $\mathbb{E}_{p(x,y)} I_{\rm EST, test|train}\le I(X,Y)$. In other words, if the set of critics that the neural network optimizes over includes the critic that saturates the DV bound, and sampling and the training algorithm are such that the globally optimal critic is found during the optimization, then the test value of the estimator will be biased down, on average. 

\paragraph*{Other approaches.}
The DEMINE and meta-DEMINE MI estimators \citep{lin2019data} also attempt to produce more efficient estimators by splitting data into validation and training data sets. The training data sets can be much larger and even use task augmentation to allow the critic to be well-learned. The data efficiency of estimating information on the validation set is decoupled from learning the critic and thus can lead to better efficiency for estimating information on the validation set. The problem with this procedure is the assumption that the training dataset is large enough for the critic to be well learned and provides no way to determine when the critic is well learned.

\paragraph*{Empirical observations.} Empirically, in all cases we reported in the paper and many others we tried but not reported, we observe that the test MI estimate is consistently biased downward (cf.~Fig.~\ref{fig:high_dim_finite_stop}). In contrast,  the train MI estimate, evaluated using the best-performing model on the test set (i.e., the model checkpoint that achieved the highest test MI before overfitting), provides a more accurate estimate of the true MI in controlled synthetic setups (e.g., cf.~Fig.~\ref{fig:high_dim_finite_vs_samples}).

\subsection{Detailed Step-by-Step Guidelines for Estimating MI}
\label{si_guideline}

Here we provide detailed description of the workflow for MI estimation,  outlined in the main text.

\paragraph*{1. Choosing an Estimator.}  
If the true MI range is unknown---which is the typical case in practice---begin with InfoNCE using a separable critic. InfoNCE is stable, low-variance, and well-behaved across a wide range of settings. Two diagnostic signals then guide any necessary switch:
\begin{itemize}[noitemsep]
    \item \textbf{InfoNCE saturates} near $\log(\text{batch size})$ (visible as a plateau in the training curve): the true MI likely exceeds this bound. Increase the batch size, or switch to SMILE or SMILE--VSIB which do not have this hard ceiling.
    \item \textbf{InfoNCE and SMILE give substantially different estimates}: this signals either a high-MI regime (where SMILE's variance dominates) or overfitting (where SMILE inflates). In both cases, SMILE--VSIB is the recommended choice, as its probabilistic regularization controls variance without imposing an artificial ceiling (see Fig.~\ref{fig:high_dim_finite_vs_samples}).
\end{itemize}
If the MI is expected to be modest and well below $\log(\text{batchsize})$, InfoNCE alone is sufficient and computationally cheaper. In linear settings (e.g., jointly Gaussian data), $I_{\rm CCA}$ is the most efficient choice and requires no training, but it is rarely the case for the data to be in this form.

\paragraph*{2. Network Design and Critic Architecture.}  
Critic architecture should match the data modality. We use MLPs for simplicity and consistency across experiments, but CNNs or transformers \citep{vaswani2017attention} may be more appropriate in other tasks. In practice, any architecture that effectively captures the statistics of the input may be used, as the critic is simply an embedding-to-MI pipeline. We have verified that CNNs also work in our setup on Noisy MNIST image data from Sec.~\ref{txt:guide} (not shown), and used a ResNet-20 backbone for the noisy CIFAR-10/100 as shown in Fig.~\ref{fig:cifar_results}.

Separable critics offer simplicity, modularity, and support for modality-specific parametrizations (e.g., using CNN for images and transformers for text), as well as faster computation via dot products. They also support variable embedding dimensionality, which plays a crucial role in practice. In contrast, concatenated critics jointly embed $(X,Y)$, capturing more complex interactions, but at the cost of more computation. They also do not provide a way to vary the latent dimensionality $k_Z$ explicitly. We recommend using concatenated critics only if the sample size is large, and computational resources are not an issue. Additionally, they should only be used if the critic dimensionality is not interesting to know, and the data modalities are homogeneous; so as to avoid mixing units. Finally, note that, in linear regimes (see Figs.~\ref{fig:low_dim_infinite_main} and \ref{fig:high_dim_infinite_main}), linear critics suffice and are significantly more efficient.

\paragraph*{3. Subsampling and Max-Test Stopping Heuristic.}  
Start with the separable critic embedding dimension of $k_Z=1$. Subsample the dataset into $\gamma$ non-overlapping subsets and evaluate the estimator on each subset. Start with $\gamma=1$ (i.e., the full dataset) and compute the MI using the max-test stopping heuristic (Fig.~\ref{fig:high_dim_finite_stop}). We typically train up to 100 epochs with an additional early stopping rule if test performance stops improving for 50 epochs. More generally, the criteria should be that the training is long enough to notice saturation or decline of the test curve. Then, for $\gamma=2$, split the dataset into two halves and compute MI on each. Continue increasing $\gamma$ up to 10 (one tenth of the full dataset in each subset). Since datasets often cannot be evenly divided, allocate $\lfloor N / \gamma \rfloor$ samples to each of the first $\gamma-1$ subsets and assign the remainder to the final subset.

\paragraph*{4. Mean and Variance Estimation.}  
For each $\gamma$, compute the sample mean $\bar{I}^{k_Z}_{\text{EST}}(\gamma)$ and standard deviation $\sigma^{k_Z}(\gamma)$ of the resulting MI values across subsets.

\paragraph*{5. Embedding Dimensionality Search.}  
Repeat steps 3 and 4 for increasing values of embedding dimensionality $k_Z$ (in separable critics). As $k_Z$ increases, MI estimates should rise until they plateau. Identify the minimal $k_Z$ beyond which estimates no longer increase significantly—this defines $k^*_Z$. Slight overestimation of $k^*_Z$ is acceptable. Note that overly large $k_Z$ may lead to severe undersampling and collapse of the information values; avoid being in this regime. Other criteria to determine $k^*_Z$ may be more suitable in specific contexts. For concatenated critics, skip this step as $k_Z$ is fixed.

\paragraph*{6. Weighted Fitting and Linear Extrapolation to $\gamma \to 0$.} \label{si_fit}
After selecting the embedding dimensionality $k^*_Z$, we estimate the MI by extrapolating the estimates taken at different data fractions to the infinite data regime, which, for a well-trained, sufficiently expressive critic, should correspond to the true MI.  We do this by fitting the curve $\bar{I}^{k_Z}_{\text{EST}}(\gamma)$ versus $\gamma$. Because we have unequal numbers of samples at each $\gamma$ (i.e., different number of subsets), a standard ordinary least squares (OLS) regression would inappropriately put too much weight to large $\gamma$. Instead, we use a weighted least squares (WLS) approach.

Traditional best practice \citep{holmes2019estimation} suggests using inverse variance across all subsets at a given data fraction as weights in WLS, with the variance smoothed via an OLS model fit to the subset variances across all fractions. This approach assumes that the estimator variance scales as $1/N$ (i.e., $\propto \gamma$), as seen with traditional estimators like $I_\text{KSG}$. However, this does not hold for neural MI estimators. We have verified that these estimators exhibit relatively uniform variance across $\gamma$ (not shown), which is likely dominated not by sample size but by stochasticity in optimization, random initializations, and inherent signal structure \citep{song2019understanding}. Thus, we instead suggest assigning weights to each subset proportional to its $1/\gamma$. For $\gamma = 1$, the single dataset is then given full weight; for $\gamma = 2$, each of the two halves is weighted by $1/2$, and so on. 

We then suggest fitting a WLS quadratic model:
\begin{equation}
\bar{I}^{k_Z}_{\text{EST}}(\gamma) = a_2 \gamma^2 + a_1 \gamma + a_0.
\end{equation}
To determine whether a linear approximation is valid, we suggest computing $\delta = |a_2/a_1|$, the relative strength of the quadratic term compared to the linear one at $\gamma=1$ (full data). If $\delta > 0.1$,  the tail of the curve (i.e., large $\gamma$) exerts too much nonlinear influence. Thus one should iteratively prune the largest $\gamma$ value and the corresponding data from the fit. One then should refit the quadratic model, recompute $\delta$, and continue removing the large  $\gamma$ values until either $\delta < 0.1$,  until a smaller range $\tilde{\gamma}$ is found, over which the information curve is approximately linear. As an additional check, after the final linear fit, one should verify that the residuals are approximately symmetric around zero across all $\gamma$ values used in the fit. A systematic negative bias in the residuals at large $\gamma$ --- even with $\delta < 0.1$ --- indicates that the linear model is inadequate and estimation should be flagged as unreliable. Such bias should also be visible in the MI vs $\gamma$ plot. Also note that the quadratic fit is a heuristic to detect the nonlinearity in the fit, and it could be mislead in some cases (cf. Fig.~\ref{si_fig:fdime_protocol} left panel), and better tests could be devised for specific cases. 

If in this process the range $\gamma\leq5$ is reached, then the information cannot be reliably extrapolated linearly. Nonlinear fits should not be used: if a term quadratic in $\gamma$ (equivalently, quadratic in $1/N$) is similar in magnitude to the linear term, then all the higher order terms are likely large too. There is no reason to truncate the Taylor series at low orders, and hence extrapolation should not be performed. The estimator should return ``estimation unreliable'' and quit. 

If, on the contrary, the reduced range $\tilde{\gamma}$ remains large (the range did not need to be reduced for any NN estimator in any of the figures in this paper), reliable estimation is possible. We suggest then to perform a final linear WLS fit over $\tilde{\gamma}$:
\begin{equation}
\bar{I}^{k_Z}_{\text{EST}}(\tilde{\gamma}) \approx b \tilde{\gamma} + c.
\end{equation}

\paragraph*{7. MI Estimation and Prediction Interval.}  
Use the intercept of the linear WLS fit of $\bar{I}^{k^*_Z}_{\text{EST}}(\tilde{\gamma})$ vs $\tilde{\gamma}$ dependence to estimate MI in the $\tilde{\gamma} = 0$ limit: $\hat{I} = c$. This provides an estimate at the infinite data limit and follows a procedure similar to \cite{strong1998entropy}. Alternatively, one can interpolate to $\gamma=1$ to find an MI for the full dataset size, less susceptible to statistical fluctuations. The corresponding prediction interval $\Delta I$ from the linear fit quantifies uncertainty in either estimate.

Note that this reported value may still be an underestimate of MI: if information is contained in features at small scales, inaccessible at the full sample size, no general purpose MI estimation algorithm will be able to recover it.

\section{Evaluations with Other Estimators and Datasets}
\label{app:evaluations}
In this section, we provide additional evaluations of our pipeline beyond the experiments reported in the main text. Specifically, we study: (i) alternative estimators in the infinite-data, low-dimensional regime, which motivated our choice of InfoNCE and SMILE; (ii) other popular estimators, including KSG \citep{kraskov2004estimating}, Sliced MI (SMI) \citep{goldfeld2021sliced}, Geodesic MI \citep{marx2022estimating}, LMI \citep{gowri2024approximating}, f-DIME \citep{letizia2024mutual}, and MINDE \citep{franzese2024minde}, in the finite-data regime of our teacher network task (Fig.~\ref{fig:high_dim_finite_vs_samples}); and (iii) evaluations on the benchmark datasets introduced in \citet{czyz2024beyond}, which allows us to compare the traditional use of InfoNCE with its modification in our pipeline.  

\subsection{Infinite-Data Regime: Motivating the Choice of InfoNCE and SMILE}
\begin{figure}[htbp]
    \centering
    \includegraphics[width=\textwidth]{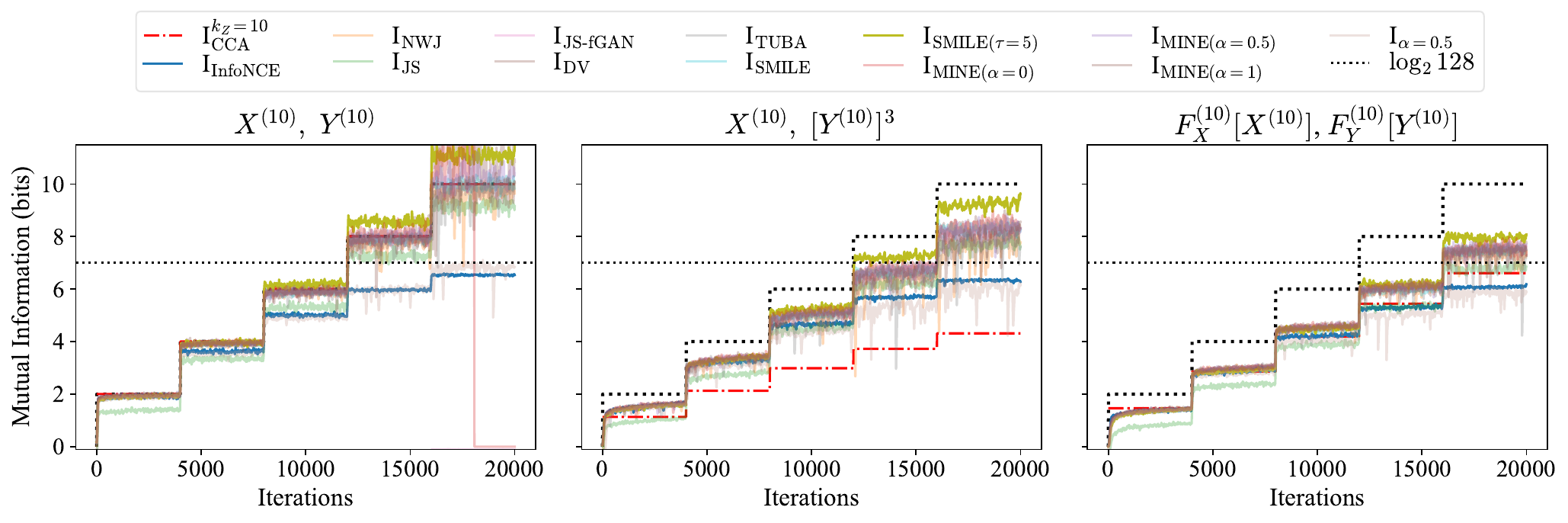}
    \caption{\textbf{Comparison of MI estimators in the low-dimensional, infinite-data regime.} We replicate the setup of Fig.~\ref{fig:low_dim_infinite_main} but include a broader selection of estimators (implementations adapted from \citet{song2019understanding, poole2019variational}). True MI is varied in discrete steps, and all estimators are run with identical batch size and sample schedules. SMILE tracks the ground truth most closely but with higher variance, while InfoNCE is biased downward at high MI yet remains stable. Other estimators either diverge during training or fail to capture the correct trends. This motivates focusing on SMILE and InfoNCE in our main study.}
    \label{si_fig:low_dim_infinite_all}
\end{figure}

See Fig.~\ref{si_fig:low_dim_infinite_all} for comparison of many different estimators and the main reason to focus on SMILE and InfoNCE in the rest of the paper.

\subsection{Finite-Data Regime: Classical and Recent Estimators}
\label{Appx:FiniteData}
\begin{figure}[htbp]
    \centering
    \includegraphics[width=0.7\textwidth]{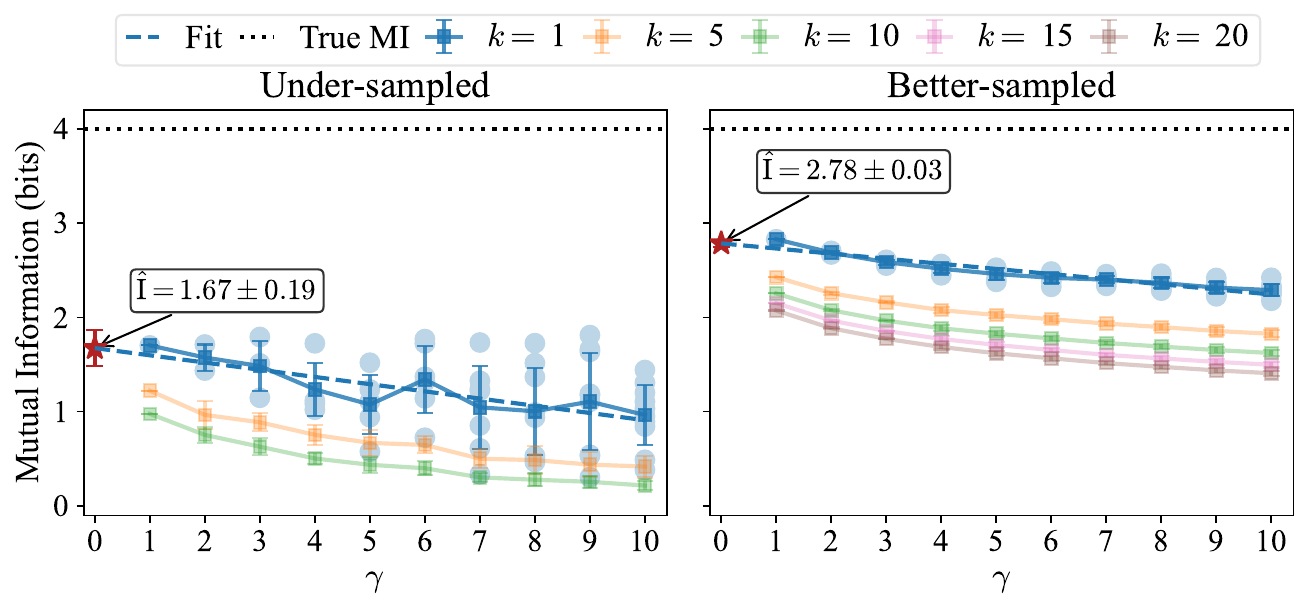}
    \caption{\textbf{Failure of the KSG estimator in high dimensions.} Estimation of $I(X;Z)$ with KSG \citep{kraskov2004estimating} on $K=500$-dimensional teacher data with latent dimension $K_Z=10$ and ground-truth MI of 4 bits. Left: undersampled ($N=256$) regime; Right: well-sampled ($N=2^{14}$). For different nearest-neighbor values $k$, MI estimates remain far below ground truth (1--2.8 bits) and do not converge under extrapolation. In contrast, neural estimators recover near-perfect values (Fig.~\ref{fig:subsampling_sufficiency}). This illustrates that KSG is unreliable for high-dimensional data, consistent with \citet{holmes2019estimation}.}
    \label{si_fig:KSG_examples}
\end{figure}

\paragraph{KSG.}
The KSG estimator \citep{kraskov2004estimating} with code adapted from  \cite{czyz2024beyond}, was used to estimate information contained in $K = 500$-dimensional data generated by teacher networks with latent dimensionality $K_Z=10$ and true MI of 4 bits, as used in Fig.~\ref{fig:subsampling_sufficiency} for neural estimators. Best practices were used to estimate the MI with KSG estimator \citep{holmes2019estimation} at different numbers of nearest neighbors $k$. Specifically, as for NN estimators, data was partitioned into $\gamma$ non-overlapping partitions and estimates for each partition were found.  The mean information and standard deviation using $\gamma$ partitions were plotted versus the number of partitions at different values of $k$ in Fig.~\ref{si_fig:KSG_examples}. The KSG estimator does not asymptotically approach the correct value of information at $\gamma=0$, corresponding to the infinite data limit, and linear extrapolations (done the same way as in Fig.~\ref{fig:subsampling_sufficiency}) is unreliable for KSG in high dimensions. Best practice is to not use the estimator in this case \citep{holmes2019estimation}. 

\paragraph{Sliced MI.}
Sliced Mutual Information (SMI) \citep{goldfeld2021sliced} does not estimate true MI but a surrogate quantity that violates the data processing inequality. While SMI may be useful for representation learning, comparing it directly to MI estimators is misleading. On our high-dimensional teacher network benchmark, available implementations yielded values far below ground truth (e.g., 1.13 bits vs.\ 4 bits), highlighting its inapplicability as a general MI estimator.

\paragraph{Geodesic MI.}
Geodesic-MI \citep{marx2022estimating} proved computationally prohibitive: runs required multiple days and $>128$GB RAM. For the better-sampled case ($N=2^{14}$), it produced estimates of 1.86 and 1.72 bits (for $k=3,10$), far below the 4-bit ground truth, showing both poor scalability and low accuracy compared to our approach.

\paragraph{LMI.}
Latent MI (LMI) estimators \citep{gowri2024approximating} assume linear-Gaussian structure and are unreliable in nonlinear settings. On our teacher network task (true MI 4 bits), LMI gave 0.65 bits (undersampled) and 1.6 bits (better-sampled), versus our $4.1\pm0.2$ and $3.7\pm0.1$. Unlike LMI’s two-step compression–estimation procedure, our end-to-end training captures nonlinear dependencies directly.

\paragraph{f-DIME.}

\begin{figure}[htbp]
\centering
\includegraphics[width=0.7\textwidth]{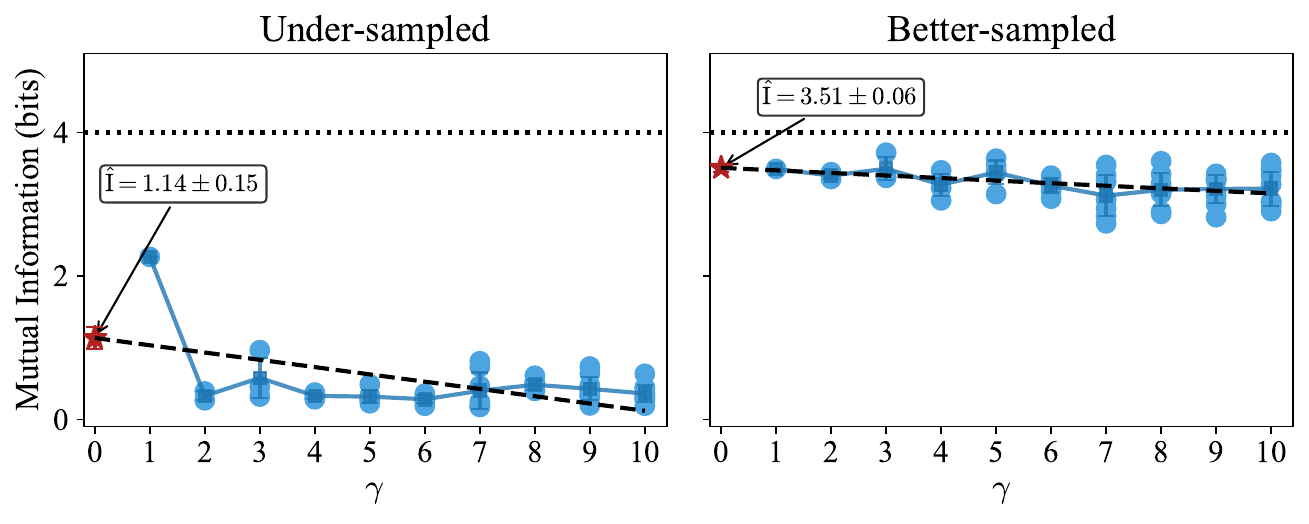}
\caption{\textbf{Our protocol applied to f-DIME (GAN-DIME variant).} We apply the subsampling-extrapolation workflow to GAN-DIME \citep{letizia2024mutual} on the teacher-network benchmark ($K=500$, $K_Z=10$, true MI 4 bits). Left: undersampled regime ($N=256$). Right: well-sampled regime ($N=2^{14}$). In the undersampled case, despite the $\delta$ criterion being formally satisfied, the points at large $\gamma$ cluster near zero while the fit extrapolates to only $1.14\pm0.15$ bits --- far below the true MI of 4 bits and with the fit line passing systematically above the data at high $\gamma$. This systematic residual bias is a visible warning sign that the estimation is unreliable, motivating an additional residual check beyond $\delta$ alone (see Appx.~\ref{si_guideline}, Step 6). In the well-sampled case, the curve is approximately linear but the extrapolated estimate ($3.51\pm0.06$ bits) falls below the true MI (4 bits), indicating a systematic underestimation bias that the protocol detects and quantifies. This demonstrates that our diagnostic workflow applies to any MI estimator, not only DV-based ones.}
\label{si_fig:fdime_protocol}
\end{figure}

We evaluated the f-DIME family of estimators \citep{letizia2024mutual} on our standard teacher-network benchmark ($K=500$, $K_Z=10$, true MI 4 bits). f-DIME addresses two known limitations of DV-based estimators: the $\log N$ saturation of contrastive methods and the partition function variance of MINE/SMILE, by decoupling the training objective from the final MI estimate. Running all three f-DIME variants (GAN-DIME, HD-DIME, KL-DIME) without modification on our benchmark, the best-performing variant (GAN-DIME) yields $3.51\pm0.05$ bits in the well-sampled regime ($N=2^{14}$) and $1.14\pm0.15$ bits undersampled ($N=256$), versus the true value of 4 bits. 
Applying our subsampling protocol to f-DIME (Appx.~Fig.~\ref{si_fig:fdime_protocol}) reveals two distinct failure modes. In the well-sampled case, the $\gamma$ curve is approximately linear but the extrapolated estimate ($3.51\pm0.06$ bits) falls systematically below the true MI (4 bits), indicating a consistent underestimation bias that the protocol detects and quantifies. In the undersampled case, the $\delta$ criterion is formally satisfied but the fit line passes systematically above the data at large $\gamma$---a visible residual bias that signals unreliable estimation even when $\delta < 0.1$. This motivates the additional residual check described in Appx.~\ref{si_guideline}, Step 6, and illustrates that our diagnostic workflow applies to any MI estimator, not only DV-based ones.

\paragraph{MINDE.}
MINDE \citep{franzese2024minde} reframes MI estimation using score-based diffusion models. Applied to our benchmark ($K=500$, $K_Z=10$, $N=2^{14}$), the MINDE-C (conditional) variant yielded 130.99 bits and the MINDE-J (joint) variant yielded 411.87 bits, against a true MI of 4 bits. The method failed entirely on the undersampled case. These results confirm that generative approaches with high model complexity require substantially more data than our target regime provides, and produce outputs that are impossible to interpret without a verification protocol.

\subsection{Standard Benchmark Suite}
\label{appx:benchmark}
\citet{czyz2024beyond} introduced a diverse suite of datasets ($1 \leq K \leq 50$) with analytically tractable MI values. They reported that InfoNCE generally outperformed alternatives (Figure 2 in their paper). We revisit these datasets (10,000 samples) using the same critic architecture as in their study (a concatenated critic with two hidden layers of width 16), but apply our bias correction, error bar estimation pipeline with held-out testing (9k train, 1k test samples).  

Table~\ref{tab:mi_results} compares ground-truth MI, results from \citet{czyz2024beyond}, and our pipeline. Grey rows mark unreliable fits ($\delta > 0.1$ or $\gamma_{\max}\leq 5$). Our method consistently matches or exceeds baseline InfoNCE performance while uniquely providing confidence intervals.  

For challenging datasets such as spirals, performance improves substantially with more expressive critics (hidden layers of width 256). Table~\ref{tab:mi_results_improved} illustrates this case. Overall, our pipeline not only matches or surpasses state-of-the-art results, but is the only one offering  error bars and internal consistency checks.

\begin{table}[htbp]
\centering
\caption{\textbf{Benchmark results on datasets from \citet{czyz2024beyond}.} True MI, results reported by \citet{czyz2024beyond} using simple InfoNCE implementation, and our pipeline with error bars. Grey rows indicate unreliable fits ($\delta > 0.1$ or $\gamma_{\max}< 5$). Our method consistently achieves estimates similar to or better than (reported in {\bf boldface}) the rivals, while uniquely reporting confidence intervals. Results reported in {\em nats} to match \cite{czyz2024beyond}.}
\label{tab:mi_results}
\footnotesize
\setlength{\tabcolsep}{10pt}
\renewcommand{\arraystretch}{1.25}
\begin{tabularx}{\textwidth}{p{7cm}|m{0.5cm}m{0.75cm}m{1.7cm}m{0.5cm}m{0.5cm}}
\toprule
Task & \makecell{True\\MI} & \makecell{simple\\InfoNCE} & \makecell{Ours} & \makecell{$\gamma_\textrm{max}$} & \makecell{$\delta$} \\
\midrule
Uniform 1 × 1 (additive noise=0.1) & 1.71 & 1.7 & 1.67 ± 0.03 & 10 & 0.041 \\
Uniform 1 × 1 (additive noise=0.75) & 0.33 & 0.3 & 0.33 ± 0.02 & 10 & 0.026 \\
\rowcolor{gray!15} Bimodal 1 × 1 & 0.41 & 0.4 & 0.39 ± 0.02 & 5 & 0.195 \\
Bivariate normal 1 × 1 & 0.41 & 0.4 & 0.39 ± 0.03 & 9 & 0.087 \\
Asinh @ Student-t 1 × 1 (dof=1) & 0.22 & 0.2 & 0.24 ± 0.02 & 9 & 0.064 \\
Asinh @ Student-t 2 × 2 (dof=1) & 0.43 & 0.4 & 0.40 ± 0.02 & 9 & 0.068 \\
Asinh @ Student-t 3 × 3 (dof=2) & 0.29 & 0.2 & \textbf{0.26 ± 0.03} & 10 & 0.098 \\
Asinh @ Student-t 5 × 5 (dof=2) & 0.45 & 0.3 & 0.34 ± 0.07 & 10 & 0.058 \\
Half-cube @ Bivariate normal 1 × 1 & 0.41 & 0.4 & 0.39 ± 0.01 & 10 & 0.077 \\
\rowcolor{gray!15} Half-cube @ Multinormal 25 × 25 (2-pair) & 1.02 & 0.8 & \textbf{1.02 ± 0.07} & 5 & 0.234 \\
Half-cube @ Multinormal 3 × 3 (2-pair) & 1.02 & 1.0 & 1.00 ± 0.02 & 10 & 0.083 \\
Half-cube @ Multinormal 5 × 5 (2-pair) & 1.02 & 1.0 & 1.01 ± 0.03 & 9 & 0.062 \\
Multinormal 2 × 2 (dense) & 0.29 & 0.3 & 0.30 ± 0.02 & 10 & 0.086 \\
Multinormal 25 × 25 (dense) & 1.29 & 1.2 & \textbf{1.26 ± 0.03} & 10 & 0.03 \\
Multinormal 3 × 3 (dense) & 0.41 & 0.4 & 0.40 ± 0.02 & 10 & 0.093 \\
Multinormal 5 × 5 (dense) & 0.59 & 0.6 & 0.60 ± 0.03 & 10 & 0.098 \\
Multinormal 50 × 50 (dense) & 1.62 & 1.4 & \textbf{1.60 ± 0.03} & 10 & 0.062 \\
Multinormal 2 × 2 (2-pair) & 1.02 & 1.0 & 1.02 ± 0.03 & 10 & 0.059 \\
\rowcolor{gray!15} Multinormal 25 × 25 (2-pair) & 1.02 & 0.9 & \textbf{0.98 ± 0.05} & 5 & 0.162 \\
\rowcolor{gray!15} Multinormal 3 × 3 (2-pair) & 1.02 & 1.0 & 1.00 ± 0.02 & 5 & 0.166 \\
Multinormal 5 × 5 (2-pair) & 1.02 & 1.0 & 1.02 ± 0.02 & 7 & 0.045 \\
Normal CDF @ Bivariate normal 1 × 1 & 0.41 & 0.4 & 0.37 ± 0.02 & 10 & 0.034 \\
Normal CDF @ Multinormal 25 × 25 (2-pair) & 1.02 & 0.8 & \textbf{0.94 ± 0.12} & 9 & 0.009 \\
Normal CDF @ Multinormal 3 × 3 (2-pair) & 1.02 & 0.9 & 0.92 ± 0.04 & 10 & 0.048 \\
Normal CDF @ Multinormal 5 × 5 (2-pair) & 1.02 & 0.9 & 0.93 ± 0.04 & 10 & 0.046 \\
Spiral @ Multinormal 25 × 25 (2-pair) & 1.02 & 0.7 & \textbf{0.97 ± 0.24} & 9 & 0.022 \\
Spiral @ Multinormal 3 × 3 (2-pair) & 1.02 & 0.6 & 0.69 ± 0.04 & 10 & 0.037 \\
Spiral @ Multinormal 5 × 5 (2-pair) & 1.02 & 0.6 & 0.64 ± 0.04 & 10 & 0.035 \\
Spiral @ Normal CDF @ Multinormal 25 × 25 (2-pair) & 1.02 & 0.8 & 0.89 ± 0.10 & 6 & 0.096 \\
Spiral @ Normal CDF @ Multinormal 3 × 3 (2-pair) & 1.02 & 0.9 & 0.89 ± 0.03 & 10 & 0.042 \\
Spiral @ Normal CDF @ Multinormal 5 × 5 (2-pair) & 1.02 & 0.9 & 0.88 ± 0.04 & 9 & 0.064 \\
\rowcolor{gray!15} Student-t 1 × 1 (dof=1) & 0.22 & 0.1 & \textbf{0.21 ± 0.03} & 5 & 0.195 \\
Student-t 2 × 2 (dof=1) & 0.43 & 0.3 & 0.27 ± 0.09 & 10 & 0.071 \\
Student-t 2 × 2 (dof=2) & 0.19 & 0.2 & 0.20 ± 0.02 & 9 & 0.054 \\
Student-t 3 × 3 (dof=2) & 0.29 & 0.2 & \textbf{0.26 ± 0.04} & 10 & 0.031 \\
Student-t 3 × 3 (dof=3) & 0.18 & 0.1 & \textbf{0.18 ± 0.03} & 9 & 0.021 \\
Student-t 5 × 5 (dof=2) & 0.45 & 0.4 & 0.37 ± 0.06 & 10 & 0.052 \\
Student-t 5 × 5 (dof=3) & 0.3 & 0.2 & 0.23 ± 0.04 & 10 & 0.054 \\
Swiss roll 2 × 1 & 0.41 & 0.4 & 0.38 ± 0.02 & 6 & 0.048 \\
\rowcolor{gray!15} Wiggly @ Bivariate normal 1 × 1 & 0.41 & 0.4 & 0.39 ± 0.01 & 5 & 0.184 \\
\bottomrule
\end{tabularx}
\end{table}

\begin{table}[htbp]
\centering
\caption{\textbf{Improved results on spiral datasets with larger critics.} For challenging cases where smaller critics underperform, increasing the hidden layer width from 16 to 256 substantially improves accuracy. Grey rows indicate unreliable fits.}
\label{tab:mi_results_improved}
\footnotesize
\setlength{\tabcolsep}{9pt}
\renewcommand{\arraystretch}{1.25}
\begin{tabularx}{\textwidth}{p{5.0cm}|m{0.5cm}m{1.0cm}m{1.5cm}m{1.8cm}m{0.4cm}m{0.4cm}}
\toprule
Task & \makecell{True\\MI} & \makecell{simple\\InfoNCE} & \makecell{Ours \\(16 hidden)} & \makecell{Ours \\(256 hidden)} & \makecell{$\gamma_\textrm{max}$} & \makecell{$\delta$} \\
\midrule
Multinormal 25 × 25 & 1.02 & 0.7 & 0.97 ± 0.24 & \textbf{1.11 ± 0.17} & 10 & 0.076 \\
Multinormal 3 × 3 & 1.02 & 0.6 & 0.69 ± 0.04 & \textbf{0.99 ± 0.05} & 10 & 0.061 \\
\rowcolor{gray!15}Multinormal 5 × 5 & 1.02 & 0.6 & 0.64 ± 0.04 & \textbf{1.04 ± 0.08} & 5 & 0.136 \\
Normal CDF @ Multinormal 25 × 25 & 1.02 & 0.8 & 0.89 ± 0.10 & \textbf{0.93 ± 0.10} & 10 & 0.067 \\
Normal CDF @ Multinormal 3 × 3 & 1.02 & 0.9 & 0.89 ± 0.03 & \textbf{1.00 ± 0.04} & 10 & 0.005 \\
Normal CDF @ Multinormal 5 × 5 & 1.02 & 0.9 & 0.88 ± 0.04 & \textbf{0.99 ± 0.03} & 10 & 0.076 \\
\bottomrule
\end{tabularx}
\end{table}

\subsection{Real-World Datasets: Noisy MNIST and Noisy CIFAR-10/100}
\label{app:real_world_datasets}

\paragraph*{Noisy MNIST.}
Each sample $(X, Y)$ consists of two $28\times28=784$-dimensional views sharing the same digit class label but drawn from distinct, non-overlapping digit instances. The $X$ view is generated by applying a random rotation (uniform in $[0, \pi/2]$) and random scaling (uniform in $[0.5, 1.5]$) to an MNIST digit \citep{Haffner1998}. The $Y$ view is formed by overlaying a different instance of the same digit class with a Perlin noise background, with noise weight drawn uniformly from $[0,1]$. Both views are normalized to $[0,1)$ intensity and flattened to 784-dimensional vectors. Since the only systematically shared information is the digit class label (10 classes), the ground-truth MI is $I(X;Y) = \log_2 10 \approx 3.3$ bits.

\paragraph*{Noisy CIFAR-10/100.}
Each sample $(X, Y)$ consists of two $32\times32$ RGB images ($K=3072$) from the same class, drawn from distinct instances and processed through different augmentation pipelines to ensure class identity is the only systematically shared information. Images are paired by cycling through class-sorted indices. View $X$ undergoes random horizontal flip ($p=0.5$), random rotation up to $15^\circ$, center crop to $32\times32$, and standard CIFAR normalization (mean $[0.491, 0.482, 0.447]$, std $[0.202, 0.199, 0.201]$). View $Y$ undergoes color jitter (brightness and contrast $\pm0.2$), random grayscale conversion (p=0.2), the same normalization, and additive Gaussian noise ($\sigma=0.05$). The theoretical upper bound on MI is $\log_2 C$ bits where $C$ is the number of classes, assuming perfectly distinct classes; in practice the true MI is lower due to visual similarity across classes. Implementation details for the backbone architecture and training procedure are in Appx.~\ref{app:implementation}.

\section{Implementation Details}
\label{app:implementation}

\subsection{General Setup}
\label{app:synthetic_and_mnist}
\paragraph*{Critic Architectures.} 
To simplify comparisons, for all synthetic experiments (i.e., Figs.~\ref{fig:low_dim_infinite_main}--\ref{fig:high_dim_finite_summarizable} and Fig.~\ref{fig:subsampling_sufficiency}) and for Noisy MNIST, we use feedforward multi-layer perceptrons (MLPs) with two hidden layers of 256 units each, initialized using Xavier uniform initialization \citep{glorot2010understanding}, and using leaky ReLU activations. For the MNIST dataset, due to the added difficulty of the task, we use deeper networks with four hidden layers of width 512. We also used CNNs for this dataset with similar results (not shown). 

Recall that our general critic has the structure $T(x, y) = f\left(g\left(x\right), h\left(y\right)\right)$.
Separable critics use one MLP for $g(x)$ and one for $h(y)$; the dot product of their outputs defines $T(x,y)$. For concatenated critics, $g$ and $h$ are identities, and $f$ is realized as an MLP  with the input $[x, y]$ (dimension $K_X + K_Y$) and output dimension 1. In both cases, the hidden width is set to the maximum of the widths used for $x$ and $y$, which are equal in our experiments.

For probabilistic critics, the architecture is identical to the deterministic case, except that the final layer is split into two heads that parameterize the mean and log-variance of a Gaussian conditional distribution of the embedding. Sampling is done via the standard reparameterization trick.

The choice of MLPs to implement the critics is just for convenience, and is not a crucial aspect of our approach. Different architectures can be used, as long as they produce a scalar critic. For example, convolutional NNs or transformers are likely to be better choice for the neural critics instead of MLPs for image or text data, respectively (see Appx.~\ref{si_guideline}).

\paragraph*{Training Details.}
All models are trained using the ADAM optimizer with a learning rate of $5 \times 10^{-4}$ and a batch size of 128, or the full dataset if it is smaller in size than 128. Each model is trained for a maximum of 100 epochs. Early stopping is applied if the test MI estimate, always evaluated on a fixed heldout batch of size 128 (even when the training set contains fewer than 128 samples), does not improve for 50 consecutive epochs or if the maximum number of epochs is reached.

\paragraph*{Other Hyperparameters.}
All unspecified hyperparameters use their default values as implemented in the \textit{PyTorch} v\texttt{2.0.1}, \textit{SciPy} v\texttt{1.11.1}, \textit{cca\_zoo} \citep{chapman2021cca} v\texttt{2.3.11}, and \textit{statsmodels} v\texttt{0.14.2} libraries.

\paragraph*{Compute Resources} 
All experiments were conducted on AWS instances. We used CPUs for $I_\text{CCA}$ and $I_\text{KSG}$, and GPUs for all neural network-based estimators. The primary instance types included \texttt{h200-8-gm1128-c192-m2048}, \texttt{a100-8-gm320-c96-m1152}, and \texttt{l40s-8-gm384-c192-m1536}.

As a reference, training a single neural estimator for one embedding dimensionality $k_Z$ on the MNIST dataset used in Fig.~\ref{fig:mnist_case_study} (the $\gamma=1$ point with $2^{14} \approx 16$k samples) takes roughly 20 seconds. Computing across all tested $k_Z$ values takes approximately 100 seconds. Subsampling experiments take a comparable amount of time; for example, evaluating two half-sized datasets (2 subsets at $\gamma=2$) takes approximately the same time as training on the full dataset. Since this procedure (as illustrated in Fig.~\ref{fig:mnist_case_study}) represents the main recommended pipeline for mutual information estimation, we report its runtime in detail. 

For completeness, other figures---such as the left panel of Fig.~\ref{fig:high_dim_finite_summarizable}---required up to 1,000 seconds for the highest sample count ($\approx 65$k), 10 trials, and multiple $k_Z$ values. The previous data point with half the number of samples ($\approx 32$k) took approximately 500 seconds.

For $I_\text{CCA}$ in Fig.~\ref{fig:high_dim_infinite_main}, computing MI at a single MI level and embedding dimension $k_Z$ took approximately 70 seconds. 
In Appx.~\ref{Appx:FiniteData}, we benchmark our approach against  other methods; so here we report their compute requirements as well. For $I_\text{KSG}$ (as used in Appx.~Fig.~\ref{si_fig:KSG_examples}), the first panel took approximately 50 seconds for the entire sweep; the second panel, which had more samples, took around 500 seconds. Both were run on CPU nodes. $I_\text{LMI}$ took about 5 minutes to run on a desktop CPU. $I_\text{Geodesic}$ took 2.25 days to run on the CPU nodes of the cluster and needed more than $128$ GB of RAM to complete. Neither of these provides estimates sample-size dependent biases or confidence intervals. That is, our approach compares favorably to others in terms of computing needs.

Overall, the total compute time, including exploratory and failed runs that led to this work, is estimated on the order of 500 compute hours across CPUs and GPUs.

\paragraph*{Smoothing and the Stopping Heuristic.} 
To avoid stopping based on high-frequency fluctuations during training, we first smooth the training and test curves using the median filter to remove outliers. We use the filter window size of 40 steps in Figs.~\ref{fig:low_dim_infinite_main} and \ref{fig:high_dim_infinite_main}, where training is done on very large number of batches/steps, and a smaller window of 5 in all other Figures.  With outliers removed, we further smooth the results with a Gaussian filter with the standard deviation $\sigma=1$. This yields the smooth test and training curves shown throughout the paper. We note that this smoothing strategy is heuristic, and other methods may be more appropriate in different settings.

\paragraph*{Final result reporting.} While it is possible to evaluate  $I_{\text{EST, train}}$ over the full training dataset, we found that evaluating it on a representative batch is sufficient for early stopping and reporting. Larger-scale averaging can be performed at evaluation time if memory allows. However, for completeness, we performed such an evaluation over the full training dataset for the task mentioned later in Appx.~\ref{appx:benchmark} and the performance is consistent, albeit with technical caveats to prevent memory related crashes.

\subsection{CIFAR-10 and CIFAR-100 Experiments}
\label{app:cifar}

\paragraph*{Backbone architecture.}
We use ResNet-20 \citep{he2016deep} adapted for $32\times32$ inputs, loaded from \citet{chenyaofo2021cifar} via \texttt{torch.hub}. The final classification head is replaced with a two-layer MLP projector: Linear($64 \to 128$)--ReLU--Linear($128 \to k_Z$), where $k_Z \in \{2, 8, 32, 128\}$ is the embedding dimension. In the \emph{frozen backbone} condition, all ResNet-20 weights are fixed and only the projector is trained. In the \emph{from-scratch} condition, the full network is randomly initialized and trained end-to-end.

\paragraph*{MI estimation.}
We use InfoNCE with a separable critic, following the same protocol as the main experiments. The Adam optimizer is used with learning rate $5\times10^{-4}$, batch size 128, maximum 100 epochs, and early stopping if the test MI does not improve for 50 consecutive epochs. A fixed held-out test batch of 128 samples is used throughout. The stopping heuristic (peak smoothed test MI, report corresponding smoothed train MI) is applied identically to the main text protocol. Sample sizes range from $N=128$ to $N=50{,}000$; embedding dimensions are $k_Z \in \{2, 8, 32, 128\}$.

\paragraph*{Other Hyperparameters, Compute Resources, Smoothing and the Stopping Heuristic, and Final result reporting.}
Other parameters and choices not mentioned here are set similar to Sec.~\ref{app:synthetic_and_mnist}.

\section{Additional Figures}
\label{app:figures}
\begin{figure}[htbp]
    \centering
    \includegraphics[width=\textwidth]{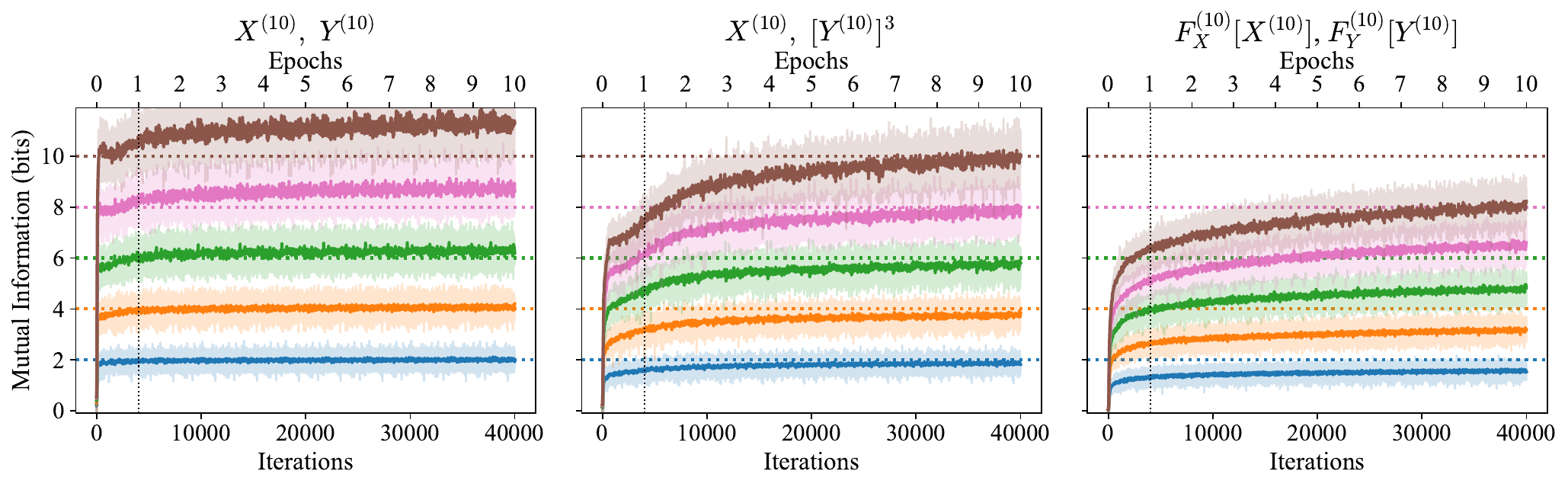}
    \caption{\textbf{Effect of continued training on MI estimation in the low-dimensional setting.} Using the SMILE estimator (with a deterministic critic), we replicate the experiment from Fig.~\ref{fig:low_dim_infinite_main}, but train for 10 epochs instead of just one, revisiting each training sample multiple times. In the raw Gaussian case (left), SMILE begins to significantly overestimate mutual information at high MI levels, consistent with its known overfitting behavior. For the $Y^3$ nonlinearly transformed case (middle), the estimator saturates to the correct MI level only after multiple epochs, suggesting that underestimation in Fig.~\ref{fig:low_dim_infinite_main} was due to insufficient training. The teacher network case (right) shows modest improvement with more training but still falls short of the true MI, reflecting the partial information loss introduced by projecting into a 10D space via non-invertible embeddings. This figure underscores the importance of the training regime and data reuse in MI estimation. While fresh batches avoid overfitting, multiple epochs can be critical for extracting information from nonlinear transformations.}
    \label{si_fig:low_dim_infinite_more_epochs}
\end{figure}


\end{document}